\documentclass[aps,showpacs,prl,reprint,superscriptaddress,longbibliography]{revtex4-1}

\usepackage[colorlinks=true,linkcolor=blue,citecolor=blue,urlcolor=blue]{hyperref}

\usepackage{hyperref}
\usepackage{setspace} 
\usepackage{graphicx}
\usepackage{amsmath}
\usepackage{color}
\usepackage{amsmath}
\usepackage{amssymb}
\usepackage{verbatim}
\usepackage{latexsym}
\usepackage{enumerate} 
\usepackage{bm} 
\usepackage[caption=false]{subfig}
\begin{document}

\title{Helium-iron compounds at terapascal pressures}

\author{Bartomeu Monserrat} \email{bm418@cam.ac.uk}
\affiliation{Department of Physics and Astronomy, Rutgers University,
  Piscataway, New Jersey 08854-8019, USA} \affiliation{TCM Group,
  Cavendish Laboratory, University of Cambridge, J.\ J.\ Thomson
  Avenue, Cambridge CB3 0HE, United Kingdom}

\author{Miguel Martinez-Canales}
\affiliation{Scottish Universities Physics Alliance, School of Physics and Astronomy, The University of Edinburgh, Edinburgh EH9 3FD, United Kingdom}
\affiliation{Centre for Science at Extreme Conditions, The University of Edinburgh, Edinburgh EH9 3FD, United Kingdom}

\author{Richard J.\ Needs}
\affiliation{TCM Group, Cavendish Laboratory, University of Cambridge,
  J.\ J.\ Thomson Avenue, Cambridge CB3 0HE, United Kingdom}

\author{Chris J.\ Pickard} 
\affiliation{Department of Materials
  Science and Metallurgy, University of Cambridge, 27 Charles Babbage
  Road, Cambridge CB3 0FS, United Kingdom}
\affiliation{Advanced Institute for Materials Research, Tohoku University
2-1-1 Katahira, Aoba, Sendai, 980-8577, Japan}

\date{\today}

\begin{abstract} 
We investigate the binary phase diagram of helium and iron using first-principles calculations. We find that helium, which is a noble gas and inert at ambient conditions, forms stable crystalline compounds with iron at terapascal pressures. A FeHe compound becomes stable above $4$~TPa, and a FeHe$_2$ compound above $12$~TPa. Melting is investigated using molecular dynamics simulations,
and a superionic phase with sublattice melting of the helium atoms is predicted. We discuss the implications of our predicted helium-iron phase diagram for interiors of giant (exo)planets and white dwarf stars. 
\end{abstract}

\pacs{}

\maketitle

Matter under extreme compression exhibits rich and unexpected behaviour, such as unconventional chemistry~\cite{Pickard_2013_ice_dissociation,Ninet_2014_nh3_ionic}, structure~\cite{WD_crust_structures}, and phases~\cite{Drozdov_2015_superconductor}. Inside planets and stars, electrons and nuclei are subject to extreme conditions of pressure and temperature, and the exploration of new physics and chemistry under these conditions is necessary for the study of astrophysical processes within the interior of the Earth~\cite{iron_melting_earth_alfe,iron_conductivity_earth_alfe,review_earth_core}, other planets~\cite{science_superionic_water,superionic_water_militzer}, or stars~\cite{Khairallah_2008_he_metal,Monserrat_2014_he_elph,McWilliams_2015_he_metal_exp}. 

Static experiments using diamond anvil cells have reached pressures of $1$~terapascal (TPa, $1$~TPa $=10^7$ atmospheres)~\cite{diamond_anvil_terapascal}, well above those at the center of the Earth but smaller than those found at the cores of giant gas planets such as Jupiter and Saturn~\cite{militzer_jupiter_model}. Higher pressures can be explored with dynamic compression experiments, as exemplified by the recent report from a team in the US National Ignition Facility that subjected diamond to pressures of $5$~TPa~\cite{diamond_nif_terapascal,pickard_news_and_views}. With high pressure experiments starting to investigate the realm of terapascal physics and chemistry, theoretical predictions are starting to emerge that reveal unexpected behaviour and complexity under these conditions. 

In this context, we use quantum mechanical calculations to explore the phase diagram of helium and iron, two of the most abundant elements in the Universe.

Helium nuclei formed in the early Universe during Big Bang nucleosynthesis, and the primordial $25$\% mass fraction of helium makes it the second most abundant element after hydrogen. In addition, thermonuclear reactions within the interiors of stars fuse hydrogen to form helium. Therefore, helium is found inside many astrophysical objects, from planets, to stars, to white dwarf stars, and it plays a central role in their behaviour. For example, recent experimental and theoretical work has shown that helium metallises at TPa pressures~\cite{Khairallah_2008_he_metal,Monserrat_2014_he_elph,McWilliams_2015_he_metal_exp}, which is higher than previously anticipated. As a consequence it has been suggested that the cooling rate of white dwarf stars is slowed by their helium-rich atmospheres, and therefore current estimates of their ages need to be revised. 

Helium has two electrons in the closed-shell $1s$ state, and is chemically inert under ambient conditions. The only known helium compounds are either metastable, involving ionised species such as HeH$_2^{+}$~\cite{Hogness_1925_heh_exp}; or are formed by weak van der Waals interactions, such as helium inside C$_{60}$~\cite{Saunders_1993_he_in_c60}. Recently, a helium-sodium compound has been reported above pressures of about $0.1$~TPa~\cite{Dong_he_na_nat_chem}. 

Iron has one of the highest binding energies per nucleon (the highest is $^{62}$Ni) and is therefore also very abundant~\cite{Burbidge_1957_stellar_nucleosynthesis_review}. It accounts for about $80$\% of the Earth's core mass~\cite{review_earth_core}, where it is found at pressures up to $0.35$~TPa, and it is responsible for the magnetic field surrounding the planet~\cite{Buffett_2000_earth_geodynamo}. Iron is not expected to exhibit magnetic order at TPa pressures, and it is predicted to occur in a series of closed-packed non-magnetic crystal structures~\cite{Pickard_2009_fe_terapascal}. Iron compounds with hydrogen, carbon, oxygen, silicon, and sulfur have been investigated at pressures of about $0.35$~TPa due to their importance for the composition of the Earth's core~\cite{earths_core_composition_review}.

In this work we investigate the possibility that, under extreme compression, helium might form stable compounds with iron. The high abundances of helium and iron make it crucial to understand the helium-iron phase diagram for astrophysical modelling of the interiors of giant planets, including the increasing number of exoplanets being discovered~\cite{Mayor_2014_exoplanets_review}, and iron-core white dwarf stars~\cite{Jordan_2012_iron_core_WD}.

Our strategy is to search for high-pressure compounds of helium and iron using first-principles quantum mechanical density functional theory (DFT) methods as implemented in the {\sc castep} code~\cite{CASTEP}, and the \textit{ab initio} random structure searching (AIRSS) method~\cite{Pickard2011}. The stability of a compound $s$ with respect to the constituent elements can be evaluated by calculating the Gibbs free energy of formation per atom $\Delta\mathcal{G}_s=(\mathcal{G}_s-(\mathcal{G}_{\mathrm{He}}N_{\mathrm{He}}+\mathcal{G}_{\mathrm{Fe}}N_{\mathrm{Fe}}))/(N_{\mathrm{He}}+N_{\mathrm{Fe}})$, where $\mathcal{G}_s$ is the Gibbs free energy of the compound $s$, $\mathcal{G}_{\mathrm{A}}$ is the Gibbs free energy per atom of A, and $N_{\mathrm{A}}$ is the number of A atoms in compound $s$. The Gibbs free energy has contributions from the electrons, which we calculate using DFT, and from the quantum and thermal nuclear motion,  which we calculate using DFT within the harmonic approximation together with the recently proposed nondiagonal supercell approach~\cite{non_diagonal} which greatly reduces the computational cost. Further details of the calculations are provided in the Supplemental Material~\cite{supp_fehe}. 

\begin{figure}
  \includegraphics[scale=0.40]{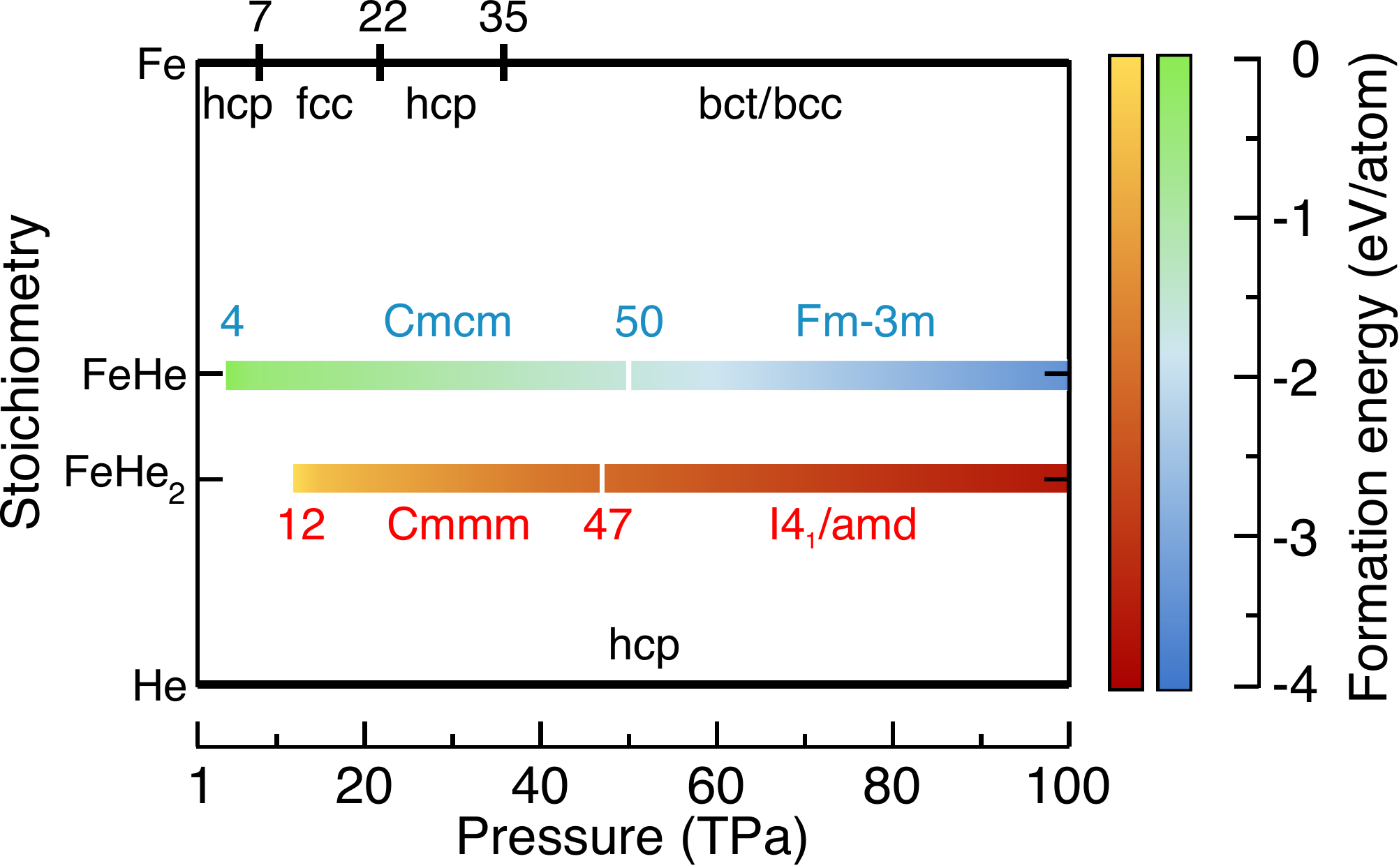}
  \caption{Pressure-composition phase diagram of the helium-iron system at the static lattice level. For FeHe, the $Cmcm$ structure is stable between $4$ and $50$~TPa, and the $Fm\overline{3}m$ structure above $50$~TPa. For FeHe$_2$, the $Cmmm$ structure is stable between $12$ and $47$~TPa, and above that pressure the stable structure is $I4_1/amd$. The formation energy per atom, calculated using $\Delta\mathcal{G}_s=(\mathcal{G}_s-(\mathcal{G}_{\mathrm{He}}N_{\mathrm{He}}+\mathcal{G}_{\mathrm{Fe}}N_{\mathrm{Fe}}))/(N_{\mathrm{He}}+N_{\mathrm{Fe}})$ as described in the text, is indicated by the gradients and approaches $-4$~eV/atom for both stoichiometries at $100$~TPa.}
      \label{fig:formation}
\end{figure}

We show the static lattice phase diagram of the helium-iron system in the pressure range $1$--$100$~TPa in Fig.~\ref{fig:formation}. Helium is predicted to adopt the hexagonal closed-packed (hcp) crystal structure at TPa pressures~\cite{Monserrat_2014_he_elph}. Iron exhibits a sequence of phase transitions at TPa pressures, starting with the hcp structure which transforms to the face-centered cubic (fcc) structure in the range $7$--$22$~TPa, then it transforms back to the hcp structure up to pressures of $35$~TPa, above which it transforms into the body-centered tetragonal (bct) structure, which approaches the body-centered cubic (bcc) structure with increasing pressure~\cite{Pickard_2009_fe_terapascal,stixrude_iron_phase_diagram}.

The structure searches find several compounds of helium and iron that are energetically competitive in the TPa pressure range, and the most stable have stoichiometries FeHe and FeHe$_2$ (see Fig.~\ref{fig:formation}). The FeHe stoichiometry first forms at $4$~TPa in a structure of orthorhombic space group $Cmcm$ containing $8$ atoms in the primitive cell, and at $50$~TPa it transforms to a $Fm\overline{3}m$ structure (rock-salt structure). The FeHe$_2$ stoichiometry appears in three distinct structures which have similar energies. The first is an orthorhombic structure of space group $Cmmm$ with nine atoms in the primitive cell, which forms around $12$~TPa. The second has a space group of $I4_1/amd$ symmetry with six atoms in the primitive cell, and becomes the most stable FeHe$_2$ structure above $47$~TPa. The third has $P6/mmm$ space group and three atoms in the primitive cell, but is not thermodynamically stable. Structure files for all of the helium-iron compounds are provided as Supplemental Material~\cite{supp_fehe}.

\begin{figure}
\subfloat[][(a) FeHe.]{
  \includegraphics[scale=0.20]{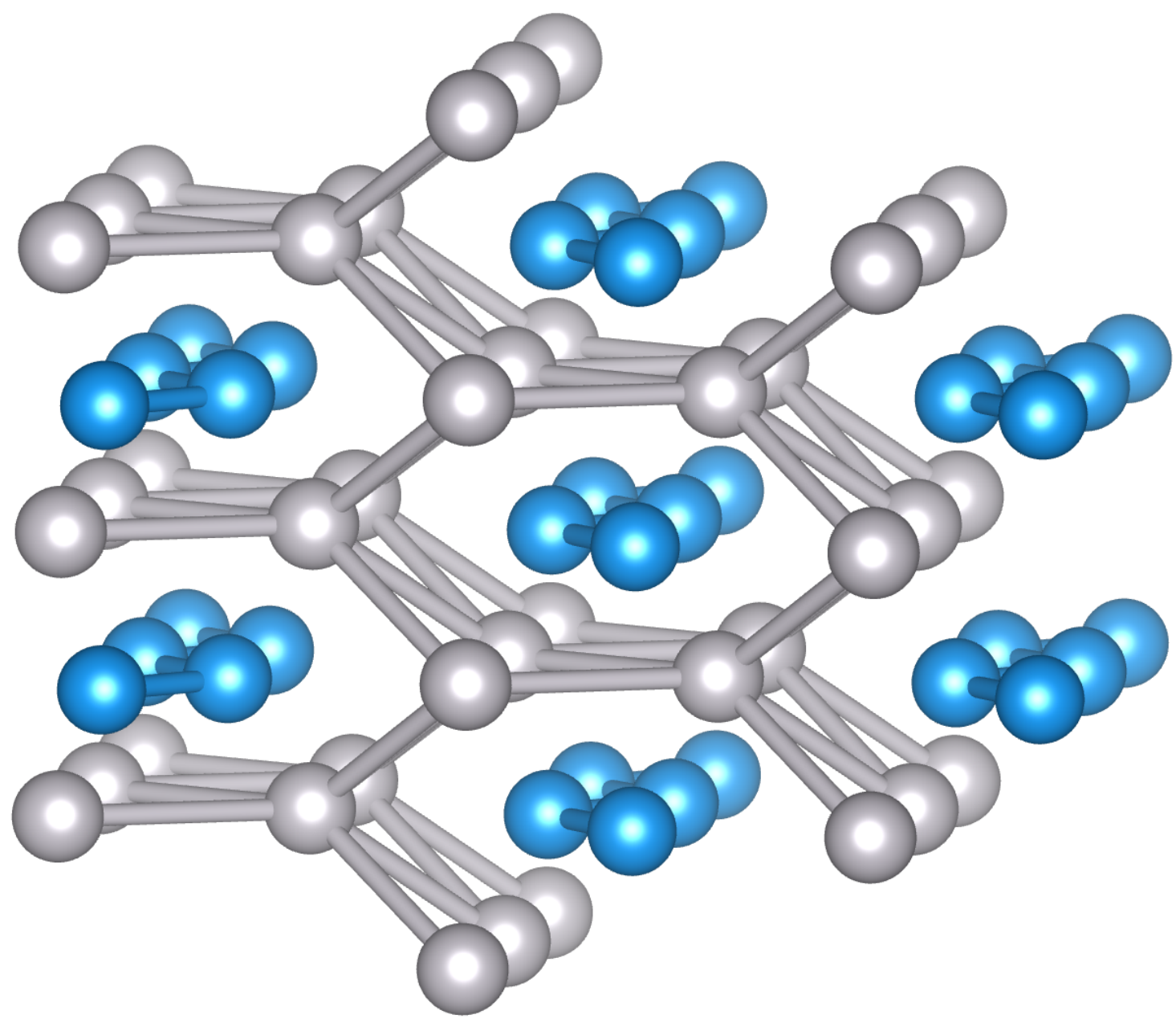}
\label{subfig:fehe}} 
\subfloat[][(b) FeHe$_2$.]{
  \includegraphics[scale=0.25]{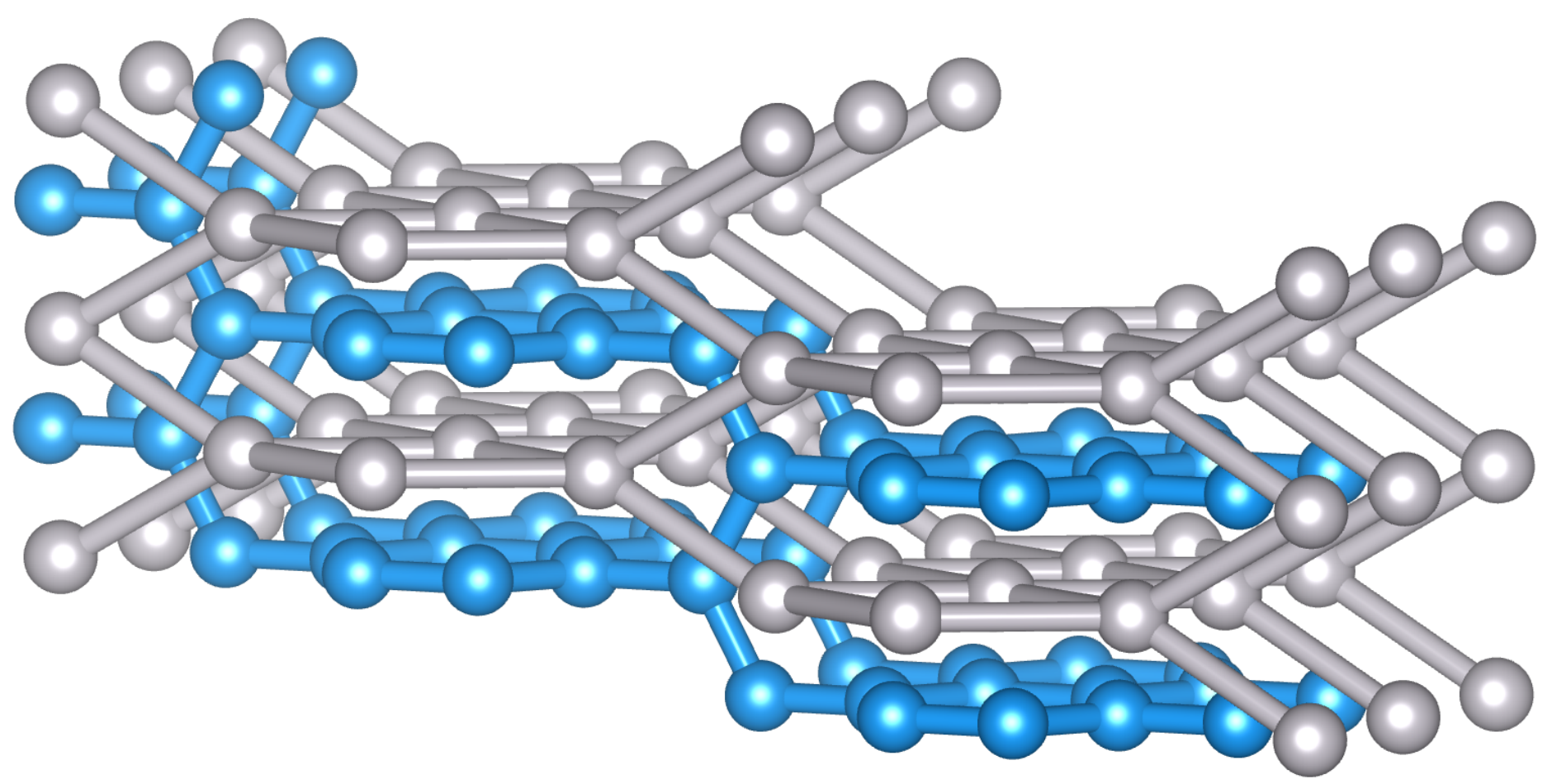}
\label{subfig:fehe2}} 
  \caption{Crystal structures of $Cmcm$ FeHe and $Cmmm$ FeHe$_2$ at $10$~TPa. Helium atoms are represented in blue, and iron atoms in grey.}
      \label{fig:structures}
\end{figure}

The helium-iron compounds that form at the lowest pressures have the FeHe $Cmcm$ and the FeHe$_2$ $Cmmm$ structures shown in Fig.~\ref{fig:structures}. The iron atoms form open channels containing helium chains in the FeHe $Cmcm$ structure (Fig.~\ref{subfig:fehe}). At $10$~TPa, the minimum He-He distance is $0.98$~\AA, the He-Fe distance is $1.16$~\AA, and the Fe-Fe distance is $1.47$~\AA. The volume per formula unit in FeHe is $2.76$~\AA$^3$, compared to $0.50$~\AA$^3$ in hcp helium and $2.27$~\AA$^3$ in both hcp and fcc iron, which add to a combined volume of $2.77$~\AA$^3$ per formula unit. In the FeHe$_2$ $Cmmm$ structure (Fig.~\ref{subfig:fehe2}), the helium atoms form hexagonal layers incorporated inside iron channels that are wider than those present in FeHe. The minimum He-He distance is $0.89$~\AA\@ at $10$~TPa, the He-Fe distance is $1.19$~\AA, and the Fe-Fe distance is larger at $1.54$~\AA. The volume per formula unit in FeHe$_2$ is $3.24$~\AA$^3$, which is smaller than that of the elements (total of $3.27$~\AA$^3$). The smaller volumes of the compounds favour their formation under pressure via the enthalpy term in the Gibbs free energy.

We next investigate the effects of temperature on the formation of helium-iron compounds upon increasing pressure. If the effects of nuclear motion are neglected, FeHe forms at pressures above $4.1$~TPa, and the inclusion of quantum and thermal nuclear motion lowers this pressure to $2.7$~TPa at $10,000$~K. FeHe$_2$ only forms at a higher pressure of about $12$~TPa, and therefore we focus on the FeHe compound to study the formation of helium-iron compounds under pressure.

We use \textit{ab initio} molecular dynamics simulations in conjunction with the Z-method~\cite{melting_z_method} to estimate the melting temperature of FeHe. These calculations are performed using the {\sc quantum espresso} package~\cite{quantum_espresso}, and the details are provided in the Supplemental Material~\cite{supp_fehe}.
The melting temperatures of helium and iron differ by thousands of degrees, suggesting that FeHe might exhibit superionicity, that is, sublattice melting of the helium component while the iron atoms oscillate around their crystallographic positions. Superionicity has been discussed before~\cite{superionic_review}, for example in a lithium-based conductor at ambient pressure~\cite{lithium_superionic_conductor}, and in the melting of ice and ammonia at extreme pressures~\cite{science_superionic_water}. Indeed, our molecular dynamics simulations demonstrate that, upon increasing temperature, the helium chains melt within the iron channels in FeHe before the iron channels themselves melt. Interestingly, metallic superionic compounds are uncommon~\cite{superionic_metallic_water}, and FeHe provides a nice platform to further investigate their properties. 

\begin{figure}
  \includegraphics[scale=0.42]{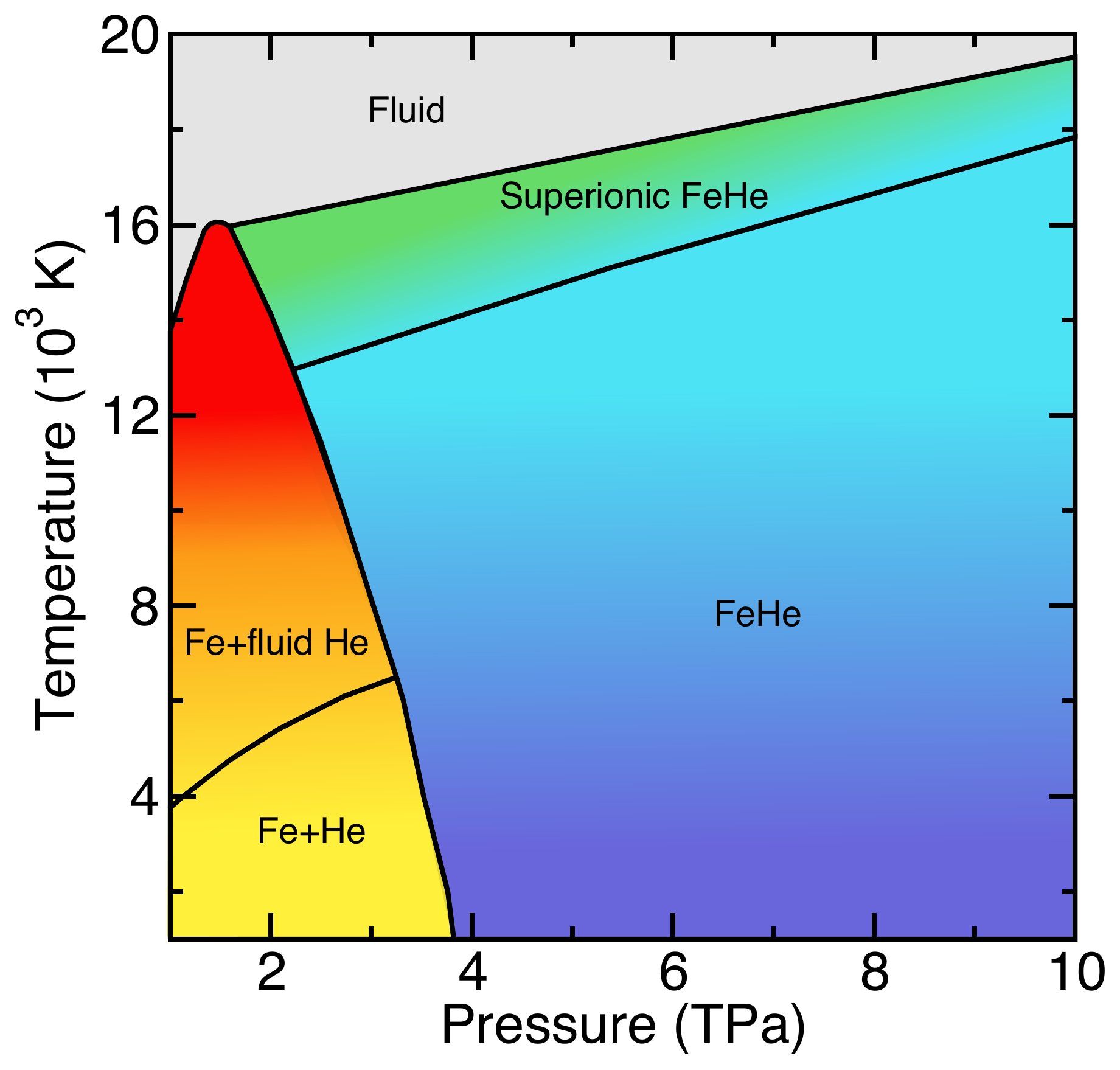}
  \caption{Helium-iron phase diagram. The solid lines indicate formation lines, sublattice melting, and full melting. The sublattice melting of FeHe occurs at $15,100\pm1,000$~K at $5.38$~TPa, and at $18,200\pm1,000$~K at $10.6$~TPa, while the full melting of FeHe occurs at $17,600\pm1,000$~K at $5.45$~TPa, and at $19,800\pm1,000$~K at $10.65$~TPa. The melting line of iron is taken from Ref.~\cite{iron_melting_dft_tpa}.} 
      \label{fig:phase_diagram}
\end{figure}

In Fig.~\ref{fig:phase_diagram} we show the proposed phase diagram for the formation of helium-iron compounds under pressures up to $10$~TPa. At low pressures, helium and iron do not mix. Below about $4,000$~K at $1$~TPa and $6,000$~K at $3$~TPa, both materials are found in the solid state, but helium melts above this temperature. Iron only melts at much higher temperatures, of the order of $15,000$~K~\cite{iron_melting_dft_tpa}. Upon increasing pressure, helium and iron form a FeHe compound between $2$ and $4$~TPa, depending on the temperature. FeHe undergoes sublattice melting of the helium atoms at temperatures between $13,000$ and $18,000$~K, depending on the pressure. The superionic phase is stable in a wide temperature and pressure range, and melting is completed at around $17,000$~K at $4$~TPa, and above $19,000$~K at $10$~TPa. 

Our results suggest that the FeHe compound should form at the pressures accessible to dynamic compression experiments. Furthermore, the formation pressure of FeHe is predicted to be within the pressure range found at the core of Jupiter, with a core-mantle boundary pressure of $4.2$~TPa and temperature of $20,000$~K, and at the highest pressures found at the centre of Saturn, with a core-mantle boundary pressure of $1$~TPa~\cite{jupiter_model_nettelmann,militzer_jupiter_model}. The interiors of exoplanets with masses similar or larger than that of Jupiter will also be subject to pressures higher than those required to form FeHe. This raises the possibility that helium is captured by iron within the interior of these planets, and potentially bound to other elements. The atmosphere of Saturn is indeed depleted of helium~\cite{saturn_atmosphere}, and the capture of helium in compounds in its interior could contribute to this phenomenon. This could also affect the helium composition of the atmospheres of giant exoplanets. White dwarf stars are subject to more extreme conditions, with helium-rich atmospheres subject to tens of terapascals, and the interiors to even higher pressures. Due to cooling, white dwarf stars exhibit temperatures in the range from only a few thousand Kelvin to hundreds of thousand of Kelvin~\cite{cool_wds}, raising the possibility that even the solid FeHe phases appear in these stars. The formation of helium compounds with other elements could alter the cooling rates of white dwarf stars, which are largely determined by the atmospheric composition, and as a consequence affect current estimates of their ages. Our results indicate that, in contrast to the inertness of helium at ambient pressure, accurate models of the composition of planets and stars should treat helium as a compound-forming element.

In conclusion, we have used first-principles methods to study the binary phase diagram of helium and iron. We have found that compounds can form at pressures of several TPa, suggesting that they might be found inside giant (exo)planets and white dwarf stars. We have also predicted that the most stable FeHe compound exhibits a superionic phase with sublattice melting of the helium atoms within a wide range of temperatures and pressures. Overall, our results show that helium can form compounds at terapascal pressures.

\acknowledgments

B.M.\ acknowledges support from the Winton Programme for the Physics of Sustainability, and from Robinson College, Cambridge, and the Cambridge
Philosophical Society for a Henslow Research Fellowship.
R.J.N.\ and C.J.P.\ acknowledge financial support from the Engineering and
Physical Sciences Research Council (EPSRC) of the UK [EP/J017639/1 and 
EP/G007489/2]. C.J.P.\ is also supported by the Royal Society through
a Royal Society Wolfson Research Merit award. The calculations were performed on the Cambridge High
Performance Computing Service facility and the Archer facility of the
UK's national high-performance computing service (for which access was
obtained via the UKCP consortium [EP/K014560/1 and EP/K013688/1]).

\bibliography{/Users/bartomeumonserrat/Documents/research/papers/references/high_pressure,/Users/bartomeumonserrat/Documents/research/papers/references/anharmonic}

\begin{thebibliography}{42}%
\makeatletter
\providecommand \@ifxundefined [1]{%
 \@ifx{#1\undefined}
}%
\providecommand \@ifnum [1]{%
 \ifnum #1\expandafter \@firstoftwo
 \else \expandafter \@secondoftwo
 \fi
}%
\providecommand \@ifx [1]{%
 \ifx #1\expandafter \@firstoftwo
 \else \expandafter \@secondoftwo
 \fi
}%
\providecommand \natexlab [1]{#1}%
\providecommand \enquote  [1]{``#1''}%
\providecommand \bibnamefont  [1]{#1}%
\providecommand \bibfnamefont [1]{#1}%
\providecommand \citenamefont [1]{#1}%
\providecommand \href@noop [0]{\@secondoftwo}%
\providecommand \href [0]{\begingroup \@sanitize@url \@href}%
\providecommand \@href[1]{\@@startlink{#1}\@@href}%
\providecommand \@@href[1]{\endgroup#1\@@endlink}%
\providecommand \@sanitize@url [0]{\catcode `\\12\catcode `\$12\catcode
  `\&12\catcode `\#12\catcode `\^12\catcode `\_12\catcode `\%12\relax}%
\providecommand \@@startlink[1]{}%
\providecommand \@@endlink[0]{}%
\providecommand \url  [0]{\begingroup\@sanitize@url \@url }%
\providecommand \@url [1]{\endgroup\@href {#1}{\urlprefix }}%
\providecommand \urlprefix  [0]{URL }%
\providecommand \Eprint [0]{\href }%
\providecommand \doibase [0]{http://dx.doi.org/}%
\providecommand \selectlanguage [0]{\@gobble}%
\providecommand \bibinfo  [0]{\@secondoftwo}%
\providecommand \bibfield  [0]{\@secondoftwo}%
\providecommand \translation [1]{[#1]}%
\providecommand \BibitemOpen [0]{}%
\providecommand \bibitemStop [0]{}%
\providecommand \bibitemNoStop [0]{.\EOS\space}%
\providecommand \EOS [0]{\spacefactor3000\relax}%
\providecommand \BibitemShut  [1]{\csname bibitem#1\endcsname}%
\let\auto@bib@innerbib\@empty
\bibitem [{\citenamefont {Pickard}\ \emph {et~al.}(2013)\citenamefont
  {Pickard}, \citenamefont {Martinez-Canales},\ and\ \citenamefont
  {Needs}}]{Pickard_2013_ice_dissociation}%
  \BibitemOpen
  \bibfield  {author} {\bibinfo {author} {\bibfnamefont {Chris~J.}\
  \bibnamefont {Pickard}}, \bibinfo {author} {\bibfnamefont {Miguel}\
  \bibnamefont {Martinez-Canales}}, \ and\ \bibinfo {author} {\bibfnamefont
  {Richard~J.}\ \bibnamefont {Needs}},\ }\bibfield  {title} {\enquote {\bibinfo
  {title} {Decomposition and terapascal phases of water ice},}\ }\href
  {http://link.aps.org/doi/10.1103/PhysRevLett.110.245701} {\bibfield
  {journal} {\bibinfo  {journal} {Phys. Rev. Lett.}\ }\textbf {\bibinfo
  {volume} {110}},\ \bibinfo {pages} {245701} (\bibinfo {year}
  {2013})}\BibitemShut {NoStop}%
\bibitem [{\citenamefont {Ninet}\ \emph {et~al.}(2014)\citenamefont {Ninet},
  \citenamefont {Datchi}, \citenamefont {Dumas}, \citenamefont {Mezouar},
  \citenamefont {Garbarino}, \citenamefont {Mafety}, \citenamefont {Pickard},
  \citenamefont {Needs},\ and\ \citenamefont {Saitta}}]{Ninet_2014_nh3_ionic}%
  \BibitemOpen
  \bibfield  {author} {\bibinfo {author} {\bibfnamefont {S.}~\bibnamefont
  {Ninet}}, \bibinfo {author} {\bibfnamefont {F.}~\bibnamefont {Datchi}},
  \bibinfo {author} {\bibfnamefont {P.}~\bibnamefont {Dumas}}, \bibinfo
  {author} {\bibfnamefont {M.}~\bibnamefont {Mezouar}}, \bibinfo {author}
  {\bibfnamefont {G.}~\bibnamefont {Garbarino}}, \bibinfo {author}
  {\bibfnamefont {A.}~\bibnamefont {Mafety}}, \bibinfo {author} {\bibfnamefont
  {C.~J.}\ \bibnamefont {Pickard}}, \bibinfo {author} {\bibfnamefont {R.~J.}\
  \bibnamefont {Needs}}, \ and\ \bibinfo {author} {\bibfnamefont {A.~M.}\
  \bibnamefont {Saitta}},\ }\bibfield  {title} {\enquote {\bibinfo {title}
  {Experimental and theoretical evidence for an ionic crystal of ammonia at
  high pressure},}\ }\href {http://link.aps.org/doi/10.1103/PhysRevB.89.174103}
  {\bibfield  {journal} {\bibinfo  {journal} {Phys. Rev. B}\ }\textbf {\bibinfo
  {volume} {89}},\ \bibinfo {pages} {174103} (\bibinfo {year}
  {2014})}\BibitemShut {NoStop}%
\bibitem [{\citenamefont {Kobyakov}\ and\ \citenamefont
  {Pethick}(2014)}]{WD_crust_structures}%
  \BibitemOpen
  \bibfield  {author} {\bibinfo {author} {\bibfnamefont {D.}~\bibnamefont
  {Kobyakov}}\ and\ \bibinfo {author} {\bibfnamefont {C.~J.}\ \bibnamefont
  {Pethick}},\ }\bibfield  {title} {\enquote {\bibinfo {title} {Towards a
  metallurgy of neutron star crusts},}\ }\href
  {http://link.aps.org/doi/10.1103/PhysRevLett.112.112504} {\bibfield
  {journal} {\bibinfo  {journal} {Phys. Rev. Lett.}\ }\textbf {\bibinfo
  {volume} {112}},\ \bibinfo {pages} {112504} (\bibinfo {year}
  {2014})}\BibitemShut {NoStop}%
\bibitem [{\citenamefont {Drozdov}\ \emph {et~al.}(2015)\citenamefont
  {Drozdov}, \citenamefont {Eremets}, \citenamefont {Troyan}, \citenamefont
  {Ksenofontov},\ and\ \citenamefont {Shylin}}]{Drozdov_2015_superconductor}%
  \BibitemOpen
  \bibfield  {author} {\bibinfo {author} {\bibfnamefont {A.~P.}\ \bibnamefont
  {Drozdov}}, \bibinfo {author} {\bibfnamefont {M.~I.}\ \bibnamefont
  {Eremets}}, \bibinfo {author} {\bibfnamefont {I.~A.}\ \bibnamefont {Troyan}},
  \bibinfo {author} {\bibfnamefont {V.}~\bibnamefont {Ksenofontov}}, \ and\
  \bibinfo {author} {\bibfnamefont {S.~I.}\ \bibnamefont {Shylin}},\ }\bibfield
   {title} {\enquote {\bibinfo {title} {Conventional superconductivity at 203
  kelvin at high pressures in the sulfur hydride system},}\ }\href
  {http://dx.doi.org/10.1038/nature14964} {\bibfield  {journal} {\bibinfo
  {journal} {Nature}\ }\textbf {\bibinfo {volume} {525}},\ \bibinfo {pages}
  {73} (\bibinfo {year} {2015})}\BibitemShut {NoStop}%
\bibitem [{\citenamefont {Alf\`{e}}\ \emph {et~al.}(1999)\citenamefont
  {Alf\`{e}}, \citenamefont {Gillan},\ and\ \citenamefont
  {Price}}]{iron_melting_earth_alfe}%
  \BibitemOpen
  \bibfield  {author} {\bibinfo {author} {\bibfnamefont {D.}~\bibnamefont
  {Alf\`{e}}}, \bibinfo {author} {\bibfnamefont {M.~J.}\ \bibnamefont
  {Gillan}}, \ and\ \bibinfo {author} {\bibfnamefont {G.~D.}\ \bibnamefont
  {Price}},\ }\bibfield  {title} {\enquote {\bibinfo {title} {The melting curve
  of iron at the pressures of the {E}arth's core from ab initio
  calculations},}\ }\href {http://dx.doi.org/10.1038/46758} {\bibfield
  {journal} {\bibinfo  {journal} {Nature}\ }\textbf {\bibinfo {volume} {401}},\
  \bibinfo {pages} {462} (\bibinfo {year} {1999})}\BibitemShut {NoStop}%
\bibitem [{\citenamefont {Pozzo}\ \emph {et~al.}(2012)\citenamefont {Pozzo},
  \citenamefont {Davies}, \citenamefont {Gubbins},\ and\ \citenamefont
  {Alf\`{e}}}]{iron_conductivity_earth_alfe}%
  \BibitemOpen
  \bibfield  {author} {\bibinfo {author} {\bibfnamefont {Monica}\ \bibnamefont
  {Pozzo}}, \bibinfo {author} {\bibfnamefont {Chris}\ \bibnamefont {Davies}},
  \bibinfo {author} {\bibfnamefont {David}\ \bibnamefont {Gubbins}}, \ and\
  \bibinfo {author} {\bibfnamefont {Dario}\ \bibnamefont {Alf\`{e}}},\
  }\bibfield  {title} {\enquote {\bibinfo {title} {Thermal and electrical
  conductivity of iron at {E}arth's core conditions},}\ }\href
  {http://dx.doi.org/10.1038/nature11031} {\bibfield  {journal} {\bibinfo
  {journal} {Nature}\ }\textbf {\bibinfo {volume} {485}},\ \bibinfo {pages}
  {355} (\bibinfo {year} {2012})}\BibitemShut {NoStop}%
\bibitem [{\citenamefont {Davies}\ \emph {et~al.}(2015)\citenamefont {Davies},
  \citenamefont {Pozzo}, \citenamefont {Gubbins},\ and\ \citenamefont
  {Alf\`{e}}}]{review_earth_core}%
  \BibitemOpen
  \bibfield  {author} {\bibinfo {author} {\bibfnamefont {Christopher}\
  \bibnamefont {Davies}}, \bibinfo {author} {\bibfnamefont {Monica}\
  \bibnamefont {Pozzo}}, \bibinfo {author} {\bibfnamefont {David}\ \bibnamefont
  {Gubbins}}, \ and\ \bibinfo {author} {\bibfnamefont {Dario}\ \bibnamefont
  {Alf\`{e}}},\ }\bibfield  {title} {\enquote {\bibinfo {title} {Constraints
  from material properties on the dynamics and evolution of {E}arth's core},}\
  }\href {http://dx.doi.org/10.1038/ngeo2492} {\bibfield  {journal} {\bibinfo
  {journal} {Nature Geosci.}\ }\textbf {\bibinfo {volume} {8}},\ \bibinfo
  {pages} {678} (\bibinfo {year} {2015})}\BibitemShut {NoStop}%
\bibitem [{\citenamefont {Cavazzoni}\ \emph {et~al.}(1999)\citenamefont
  {Cavazzoni}, \citenamefont {Chiarotti}, \citenamefont {Scandolo},
  \citenamefont {Tosatti}, \citenamefont {Bernasconi},\ and\ \citenamefont
  {Parrinello}}]{science_superionic_water}%
  \BibitemOpen
  \bibfield  {author} {\bibinfo {author} {\bibfnamefont {C.}~\bibnamefont
  {Cavazzoni}}, \bibinfo {author} {\bibfnamefont {G.~L.}\ \bibnamefont
  {Chiarotti}}, \bibinfo {author} {\bibfnamefont {S.}~\bibnamefont {Scandolo}},
  \bibinfo {author} {\bibfnamefont {E.}~\bibnamefont {Tosatti}}, \bibinfo
  {author} {\bibfnamefont {M.}~\bibnamefont {Bernasconi}}, \ and\ \bibinfo
  {author} {\bibfnamefont {M.}~\bibnamefont {Parrinello}},\ }\bibfield  {title}
  {\enquote {\bibinfo {title} {Superionic and metallic states of water and
  ammonia at giant planet conditions},}\ }\href
  {http://science.sciencemag.org/content/283/5398/44} {\bibfield  {journal}
  {\bibinfo  {journal} {Science}\ }\textbf {\bibinfo {volume} {283}},\ \bibinfo
  {pages} {44--46} (\bibinfo {year} {1999})}\BibitemShut {NoStop}%
\bibitem [{\citenamefont {Wilson}\ \emph {et~al.}(2013)\citenamefont {Wilson},
  \citenamefont {Wong},\ and\ \citenamefont
  {Militzer}}]{superionic_water_militzer}%
  \BibitemOpen
  \bibfield  {author} {\bibinfo {author} {\bibfnamefont {Hugh~F.}\ \bibnamefont
  {Wilson}}, \bibinfo {author} {\bibfnamefont {Michael~L.}\ \bibnamefont
  {Wong}}, \ and\ \bibinfo {author} {\bibfnamefont {Burkhard}\ \bibnamefont
  {Militzer}},\ }\bibfield  {title} {\enquote {\bibinfo {title} {Superionic to
  superionic phase change in water: Consequences for the interiors of {U}ranus
  and {N}eptune},}\ }\href
  {http://link.aps.org/doi/10.1103/PhysRevLett.110.151102} {\bibfield
  {journal} {\bibinfo  {journal} {Phys. Rev. Lett.}\ }\textbf {\bibinfo
  {volume} {110}},\ \bibinfo {pages} {151102} (\bibinfo {year}
  {2013})}\BibitemShut {NoStop}%
\bibitem [{\citenamefont {Khairallah}\ and\ \citenamefont
  {Militzer}(2008)}]{Khairallah_2008_he_metal}%
  \BibitemOpen
  \bibfield  {author} {\bibinfo {author} {\bibfnamefont {S.~A.}\ \bibnamefont
  {Khairallah}}\ and\ \bibinfo {author} {\bibfnamefont {B.}~\bibnamefont
  {Militzer}},\ }\bibfield  {title} {\enquote {\bibinfo {title}
  {First-principles studies of the metallization and the equation of state of
  solid helium},}\ }\href
  {http://link.aps.org/doi/10.1103/PhysRevLett.101.106407} {\bibfield
  {journal} {\bibinfo  {journal} {Phys. Rev. Lett.}\ }\textbf {\bibinfo
  {volume} {101}},\ \bibinfo {pages} {106407} (\bibinfo {year}
  {2008})}\BibitemShut {NoStop}%
\bibitem [{\citenamefont {Monserrat}\ \emph {et~al.}(2014)\citenamefont
  {Monserrat}, \citenamefont {Drummond}, \citenamefont {Pickard},\ and\
  \citenamefont {Needs}}]{Monserrat_2014_he_elph}%
  \BibitemOpen
  \bibfield  {author} {\bibinfo {author} {\bibfnamefont {Bartomeu}\
  \bibnamefont {Monserrat}}, \bibinfo {author} {\bibfnamefont {N.~D.}\
  \bibnamefont {Drummond}}, \bibinfo {author} {\bibfnamefont {Chris~J.}\
  \bibnamefont {Pickard}}, \ and\ \bibinfo {author} {\bibfnamefont {R.~J.}\
  \bibnamefont {Needs}},\ }\bibfield  {title} {\enquote {\bibinfo {title}
  {Electron-phonon coupling and the metallization of solid helium at terapascal
  pressures},}\ }\href {http://link.aps.org/doi/10.1103/PhysRevLett.112.055504}
  {\bibfield  {journal} {\bibinfo  {journal} {Phys. Rev. Lett.}\ }\textbf
  {\bibinfo {volume} {112}},\ \bibinfo {pages} {055504} (\bibinfo {year}
  {2014})}\BibitemShut {NoStop}%
\bibitem [{\citenamefont {McWilliams}\ \emph {et~al.}(2015)\citenamefont
  {McWilliams}, \citenamefont {Dalton}, \citenamefont {Konôpková},
  \citenamefont {Mahmood},\ and\ \citenamefont
  {Goncharov}}]{McWilliams_2015_he_metal_exp}%
  \BibitemOpen
  \bibfield  {author} {\bibinfo {author} {\bibfnamefont {R.~Stewart}\
  \bibnamefont {McWilliams}}, \bibinfo {author} {\bibfnamefont {D.~Allen}\
  \bibnamefont {Dalton}}, \bibinfo {author} {\bibfnamefont {Zuzana}\
  \bibnamefont {Konôpková}}, \bibinfo {author} {\bibfnamefont {Mohammad~F.}\
  \bibnamefont {Mahmood}}, \ and\ \bibinfo {author} {\bibfnamefont
  {Alexander~F.}\ \bibnamefont {Goncharov}},\ }\bibfield  {title} {\enquote
  {\bibinfo {title} {Opacity and conductivity measurements in noble gases at
  conditions of planetary and stellar interiors},}\ }\href
  {http://www.pnas.org/content/112/26/7925.abstract} {\bibfield  {journal}
  {\bibinfo  {journal} {Proc. Natl. Acad. Sci. USA}\ }\textbf {\bibinfo
  {volume} {112}},\ \bibinfo {pages} {7925--7930} (\bibinfo {year}
  {2015})}\BibitemShut {NoStop}%
\bibitem [{\citenamefont {Dubrovinskaia}\ \emph {et~al.}(2016)\citenamefont
  {Dubrovinskaia}, \citenamefont {Dubrovinsky}, \citenamefont {Solopova},
  \citenamefont {Abakumov}, \citenamefont {Turner}, \citenamefont {Hanfland},
  \citenamefont {Bykova}, \citenamefont {Bykov}, \citenamefont {Prescher},
  \citenamefont {Prakapenka}, \citenamefont {Petitgirard}, \citenamefont
  {Chuvashova}, \citenamefont {Gasharova}, \citenamefont {Mathis},
  \citenamefont {Ershov}, \citenamefont {Snigireva},\ and\ \citenamefont
  {Snigirev}}]{diamond_anvil_terapascal}%
  \BibitemOpen
  \bibfield  {author} {\bibinfo {author} {\bibfnamefont {Natalia}\ \bibnamefont
  {Dubrovinskaia}}, \bibinfo {author} {\bibfnamefont {Leonid}\ \bibnamefont
  {Dubrovinsky}}, \bibinfo {author} {\bibfnamefont {Natalia~A.}\ \bibnamefont
  {Solopova}}, \bibinfo {author} {\bibfnamefont {Artem}\ \bibnamefont
  {Abakumov}}, \bibinfo {author} {\bibfnamefont {Stuart}\ \bibnamefont
  {Turner}}, \bibinfo {author} {\bibfnamefont {Michael}\ \bibnamefont
  {Hanfland}}, \bibinfo {author} {\bibfnamefont {Elena}\ \bibnamefont
  {Bykova}}, \bibinfo {author} {\bibfnamefont {Maxim}\ \bibnamefont {Bykov}},
  \bibinfo {author} {\bibfnamefont {Clemens}\ \bibnamefont {Prescher}},
  \bibinfo {author} {\bibfnamefont {Vitali~B.}\ \bibnamefont {Prakapenka}},
  \bibinfo {author} {\bibfnamefont {Sylvain}\ \bibnamefont {Petitgirard}},
  \bibinfo {author} {\bibfnamefont {Irina}\ \bibnamefont {Chuvashova}},
  \bibinfo {author} {\bibfnamefont {Biliana}\ \bibnamefont {Gasharova}},
  \bibinfo {author} {\bibfnamefont {Yves-Laurent}\ \bibnamefont {Mathis}},
  \bibinfo {author} {\bibfnamefont {Petr}\ \bibnamefont {Ershov}}, \bibinfo
  {author} {\bibfnamefont {Irina}\ \bibnamefont {Snigireva}}, \ and\ \bibinfo
  {author} {\bibfnamefont {Anatoly}\ \bibnamefont {Snigirev}},\ }\bibfield
  {title} {\enquote {\bibinfo {title} {Terapascal static pressure generation
  with ultrahigh yield strength nanodiamond},}\ }\href
  {http://advances.sciencemag.org/content/2/7/e1600341} {\bibfield  {journal}
  {\bibinfo  {journal} {Sci. Adv.}\ }\textbf {\bibinfo {volume} {2}} (\bibinfo
  {year} {2016})}\BibitemShut {NoStop}%
\bibitem [{\citenamefont {Hubbard}\ and\ \citenamefont
  {Militzer}(2016)}]{militzer_jupiter_model}%
  \BibitemOpen
  \bibfield  {author} {\bibinfo {author} {\bibfnamefont {W.~B.}\ \bibnamefont
  {Hubbard}}\ and\ \bibinfo {author} {\bibfnamefont {B.}~\bibnamefont
  {Militzer}},\ }\bibfield  {title} {\enquote {\bibinfo {title} {A preliminary
  {J}upiter model},}\ }\href {http://stacks.iop.org/0004-637X/820/i=1/a=80}
  {\bibfield  {journal} {\bibinfo  {journal} {Astrophys. J.}\ }\textbf
  {\bibinfo {volume} {820}},\ \bibinfo {pages} {80} (\bibinfo {year}
  {2016})}\BibitemShut {NoStop}%
\bibitem [{\citenamefont {Smith}\ \emph {et~al.}(2014)\citenamefont {Smith},
  \citenamefont {Eggert}, \citenamefont {Jeanloz}, \citenamefont {Duffy},
  \citenamefont {Braun}, \citenamefont {Patterson}, \citenamefont {Rudd},
  \citenamefont {Biener}, \citenamefont {Lazicki}, \citenamefont {Hamza},
  \citenamefont {Wang}, \citenamefont {Braun}, \citenamefont {Benedict},
  \citenamefont {Celliers},\ and\ \citenamefont
  {Collins}}]{diamond_nif_terapascal}%
  \BibitemOpen
  \bibfield  {author} {\bibinfo {author} {\bibfnamefont {R.~F.}\ \bibnamefont
  {Smith}}, \bibinfo {author} {\bibfnamefont {J.~H.}\ \bibnamefont {Eggert}},
  \bibinfo {author} {\bibfnamefont {R.}~\bibnamefont {Jeanloz}}, \bibinfo
  {author} {\bibfnamefont {T.~S.}\ \bibnamefont {Duffy}}, \bibinfo {author}
  {\bibfnamefont {D.~G.}\ \bibnamefont {Braun}}, \bibinfo {author}
  {\bibfnamefont {J.~R.}\ \bibnamefont {Patterson}}, \bibinfo {author}
  {\bibfnamefont {R.~E.}\ \bibnamefont {Rudd}}, \bibinfo {author}
  {\bibfnamefont {J.}~\bibnamefont {Biener}}, \bibinfo {author} {\bibfnamefont
  {A.~E.}\ \bibnamefont {Lazicki}}, \bibinfo {author} {\bibfnamefont {A.~V.}\
  \bibnamefont {Hamza}}, \bibinfo {author} {\bibfnamefont {J.}~\bibnamefont
  {Wang}}, \bibinfo {author} {\bibfnamefont {T.}~\bibnamefont {Braun}},
  \bibinfo {author} {\bibfnamefont {L.~X.}\ \bibnamefont {Benedict}}, \bibinfo
  {author} {\bibfnamefont {P.~M.}\ \bibnamefont {Celliers}}, \ and\ \bibinfo
  {author} {\bibfnamefont {G.~W.}\ \bibnamefont {Collins}},\ }\bibfield
  {title} {\enquote {\bibinfo {title} {Ramp compression of diamond to five
  terapascals},}\ }\href {http://dx.doi.org/10.1038/nature13526} {\bibfield
  {journal} {\bibinfo  {journal} {Nature}\ }\textbf {\bibinfo {volume} {511}},\
  \bibinfo {pages} {330} (\bibinfo {year} {2014})}\BibitemShut {NoStop}%
\bibitem [{\citenamefont {Pickard}\ and\ \citenamefont
  {Needs}(2014)}]{pickard_news_and_views}%
  \BibitemOpen
  \bibfield  {author} {\bibinfo {author} {\bibfnamefont {Chris~J.}\
  \bibnamefont {Pickard}}\ and\ \bibinfo {author} {\bibfnamefont {Richard~J.}\
  \bibnamefont {Needs}},\ }\bibfield  {title} {\enquote {\bibinfo {title}
  {High-pressure physics: Piling on the pressure},}\ }\href
  {http://dx.doi.org/10.1038/511294a} {\bibfield  {journal} {\bibinfo
  {journal} {Nature}\ }\textbf {\bibinfo {volume} {511}},\ \bibinfo {pages}
  {294} (\bibinfo {year} {2014})}\BibitemShut {NoStop}%
\bibitem [{\citenamefont {Hogness}\ and\ \citenamefont
  {Lunn}(1925)}]{Hogness_1925_heh_exp}%
  \BibitemOpen
  \bibfield  {author} {\bibinfo {author} {\bibfnamefont {T.~R.}\ \bibnamefont
  {Hogness}}\ and\ \bibinfo {author} {\bibfnamefont {E.~G.}\ \bibnamefont
  {Lunn}},\ }\bibfield  {title} {\enquote {\bibinfo {title} {The ionization of
  hydrogen by electron impact as interpreted by positive ray analysis},}\
  }\href {http://link.aps.org/doi/10.1103/PhysRev.26.44} {\bibfield  {journal}
  {\bibinfo  {journal} {Phys. Rev.}\ }\textbf {\bibinfo {volume} {26}},\
  \bibinfo {pages} {44--55} (\bibinfo {year} {1925})}\BibitemShut {NoStop}%
\bibitem [{\citenamefont {Saunders}\ \emph {et~al.}(1993)\citenamefont
  {Saunders}, \citenamefont {Jim\'{e}nez-V\'{a}zquez}, \citenamefont {Cross},\
  and\ \citenamefont {Poreda}}]{Saunders_1993_he_in_c60}%
  \BibitemOpen
  \bibfield  {author} {\bibinfo {author} {\bibfnamefont {Martin}\ \bibnamefont
  {Saunders}}, \bibinfo {author} {\bibfnamefont {Hugo~A.}\ \bibnamefont
  {Jim\'{e}nez-V\'{a}zquez}}, \bibinfo {author} {\bibfnamefont {R.~James}\
  \bibnamefont {Cross}}, \ and\ \bibinfo {author} {\bibfnamefont {Robert~J.}\
  \bibnamefont {Poreda}},\ }\bibfield  {title} {\enquote {\bibinfo {title}
  {Stable compounds of helium and neon: {H}e@{C}60 and {N}e@{C}60},}\ }\href
  {http://www.sciencemag.org/content/259/5100/1428.abstract} {\bibfield
  {journal} {\bibinfo  {journal} {Science}\ }\textbf {\bibinfo {volume}
  {259}},\ \bibinfo {pages} {1428--1430} (\bibinfo {year} {1993})}\BibitemShut
  {NoStop}%
\bibitem [{\citenamefont {Dong}\ \emph {et~al.}(2017)\citenamefont {Dong},
  \citenamefont {Oganov}, \citenamefont {Goncharov}, \citenamefont {Stavrou},
  \citenamefont {Lobanov}, \citenamefont {Saleh}, \citenamefont {Qian},
  \citenamefont {Zhu}, \citenamefont {Gatti}, \citenamefont {Deringer},
  \citenamefont {Dronskowski}, \citenamefont {Zhou}, \citenamefont
  {Prakapenka}, \citenamefont {Kon\^{o}pkov\'{a}}, \citenamefont {Popov},
  \citenamefont {Boldyrev},\ and\ \citenamefont {Wang}}]{Dong_he_na_nat_chem}%
  \BibitemOpen
  \bibfield  {author} {\bibinfo {author} {\bibfnamefont {Xiao}\ \bibnamefont
  {Dong}}, \bibinfo {author} {\bibfnamefont {Artem~R.}\ \bibnamefont {Oganov}},
  \bibinfo {author} {\bibfnamefont {Alexander~F.}\ \bibnamefont {Goncharov}},
  \bibinfo {author} {\bibfnamefont {Elissaios}\ \bibnamefont {Stavrou}},
  \bibinfo {author} {\bibfnamefont {Sergey}\ \bibnamefont {Lobanov}}, \bibinfo
  {author} {\bibfnamefont {Gabriele}\ \bibnamefont {Saleh}}, \bibinfo {author}
  {\bibfnamefont {Guang-Rui}\ \bibnamefont {Qian}}, \bibinfo {author}
  {\bibfnamefont {Qiang}\ \bibnamefont {Zhu}}, \bibinfo {author} {\bibfnamefont
  {Carlo}\ \bibnamefont {Gatti}}, \bibinfo {author} {\bibfnamefont {Volker~L.}\
  \bibnamefont {Deringer}}, \bibinfo {author} {\bibfnamefont {Richard}\
  \bibnamefont {Dronskowski}}, \bibinfo {author} {\bibfnamefont {Xiang-Feng}\
  \bibnamefont {Zhou}}, \bibinfo {author} {\bibfnamefont {Vitali~B.}\
  \bibnamefont {Prakapenka}}, \bibinfo {author} {\bibfnamefont
  {Z.}~\bibnamefont {Kon\^{o}pkov\'{a}}}, \bibinfo {author} {\bibfnamefont
  {Ivan~A.}\ \bibnamefont {Popov}}, \bibinfo {author} {\bibfnamefont
  {Alexander~I.}\ \bibnamefont {Boldyrev}}, \ and\ \bibinfo {author}
  {\bibfnamefont {Hui-Tian}\ \bibnamefont {Wang}},\ }\bibfield  {title}
  {\enquote {\bibinfo {title} {A stable compound of helium and sodium at high
  pressure},}\ }\href {http://dx.doi.org/10.1038/nchem.2716} {\bibfield
  {journal} {\bibinfo  {journal} {Nat. Chem.}\ }\textbf {\bibinfo {volume}
  {9}},\ \bibinfo {pages} {440} (\bibinfo {year} {2017})}\BibitemShut {NoStop}%
\bibitem [{\citenamefont {Burbidge}\ \emph {et~al.}(1957)\citenamefont
  {Burbidge}, \citenamefont {Burbidge}, \citenamefont {Fowler},\ and\
  \citenamefont {Hoyle}}]{Burbidge_1957_stellar_nucleosynthesis_review}%
  \BibitemOpen
  \bibfield  {author} {\bibinfo {author} {\bibfnamefont {E.~Margaret}\
  \bibnamefont {Burbidge}}, \bibinfo {author} {\bibfnamefont {G.~R.}\
  \bibnamefont {Burbidge}}, \bibinfo {author} {\bibfnamefont {William~A.}\
  \bibnamefont {Fowler}}, \ and\ \bibinfo {author} {\bibfnamefont
  {F.}~\bibnamefont {Hoyle}},\ }\bibfield  {title} {\enquote {\bibinfo {title}
  {Synthesis of the elements in stars},}\ }\href
  {http://link.aps.org/doi/10.1103/RevModPhys.29.547} {\bibfield  {journal}
  {\bibinfo  {journal} {Rev. Mod. Phys.}\ }\textbf {\bibinfo {volume} {29}},\
  \bibinfo {pages} {547--650} (\bibinfo {year} {1957})}\BibitemShut {NoStop}%
\bibitem [{\citenamefont {Buffett}(2000)}]{Buffett_2000_earth_geodynamo}%
  \BibitemOpen
  \bibfield  {author} {\bibinfo {author} {\bibfnamefont {Bruce~A.}\
  \bibnamefont {Buffett}},\ }\bibfield  {title} {\enquote {\bibinfo {title}
  {Earth's core and the geodynamo},}\ }\href
  {http://www.sciencemag.org/content/288/5473/2007.abstract} {\bibfield
  {journal} {\bibinfo  {journal} {Science}\ }\textbf {\bibinfo {volume}
  {288}},\ \bibinfo {pages} {2007--2012} (\bibinfo {year} {2000})}\BibitemShut
  {NoStop}%
\bibitem [{\citenamefont {Pickard}\ and\ \citenamefont
  {Needs}(2009)}]{Pickard_2009_fe_terapascal}%
  \BibitemOpen
  \bibfield  {author} {\bibinfo {author} {\bibfnamefont {C.~J.}\ \bibnamefont
  {Pickard}}\ and\ \bibinfo {author} {\bibfnamefont {R.~J.}\ \bibnamefont
  {Needs}},\ }\bibfield  {title} {\enquote {\bibinfo {title} {Stable phases of
  iron at terapascal pressures},}\ }\href
  {http://stacks.iop.org/0953-8984/21/i=45/a=452205} {\bibfield  {journal}
  {\bibinfo  {journal} {J. Phys. Condens. Matter}\ }\textbf {\bibinfo {volume}
  {21}},\ \bibinfo {pages} {452205} (\bibinfo {year} {2009})}\BibitemShut
  {NoStop}%
\bibitem [{\citenamefont {Poirier}(1994)}]{earths_core_composition_review}%
  \BibitemOpen
  \bibfield  {author} {\bibinfo {author} {\bibfnamefont {Jean-Paul}\
  \bibnamefont {Poirier}},\ }\bibfield  {title} {\enquote {\bibinfo {title}
  {Light elements in the {E}arth's outer core: A critical review},}\ }\href
  {http://www.sciencedirect.com/science/article/pii/0031920194901201}
  {\bibfield  {journal} {\bibinfo  {journal} {Phys. Earth Planet. Inter.}\
  }\textbf {\bibinfo {volume} {85}},\ \bibinfo {pages} {319 -- 337} (\bibinfo
  {year} {1994})}\BibitemShut {NoStop}%
\bibitem [{\citenamefont {Mayor}\ \emph {et~al.}(2014)\citenamefont {Mayor},
  \citenamefont {Lovis},\ and\ \citenamefont
  {Santos}}]{Mayor_2014_exoplanets_review}%
  \BibitemOpen
  \bibfield  {author} {\bibinfo {author} {\bibfnamefont {Michel}\ \bibnamefont
  {Mayor}}, \bibinfo {author} {\bibfnamefont {Christophe}\ \bibnamefont
  {Lovis}}, \ and\ \bibinfo {author} {\bibfnamefont {Nuno~C.}\ \bibnamefont
  {Santos}},\ }\bibfield  {title} {\enquote {\bibinfo {title} {Doppler
  spectroscopy as a path to the detection of {E}arth-like planets},}\ }\href
  {http://dx.doi.org/10.1038/nature13780} {\bibfield  {journal} {\bibinfo
  {journal} {Nature}\ }\textbf {\bibinfo {volume} {513}},\ \bibinfo {pages}
  {328} (\bibinfo {year} {2014})}\BibitemShut {NoStop}%
\bibitem [{\citenamefont {IV}\ \emph {et~al.}(2012)\citenamefont {IV},
  \citenamefont {Perets}, \citenamefont {Fisher},\ and\ \citenamefont {van
  Rossum}}]{Jordan_2012_iron_core_WD}%
  \BibitemOpen
  \bibfield  {author} {\bibinfo {author} {\bibfnamefont {George C.~Jordan}\
  \bibnamefont {IV}}, \bibinfo {author} {\bibfnamefont {Hagai~B.}\ \bibnamefont
  {Perets}}, \bibinfo {author} {\bibfnamefont {Robert~T.}\ \bibnamefont
  {Fisher}}, \ and\ \bibinfo {author} {\bibfnamefont {Daniel~R.}\ \bibnamefont
  {van Rossum}},\ }\bibfield  {title} {\enquote {\bibinfo {title}
  {Failed-detonation supernovae: Subluminous low-velocity {I}a supernovae and
  their kicked remnant white dwarfs with iron-rich cores},}\ }\href
  {http://stacks.iop.org/2041-8205/761/i=2/a=L23} {\bibfield  {journal}
  {\bibinfo  {journal} {The Astrophysical Journal Letters}\ }\textbf {\bibinfo
  {volume} {761}},\ \bibinfo {pages} {L23} (\bibinfo {year}
  {2012})}\BibitemShut {NoStop}%
\bibitem [{\citenamefont {Clark}\ \emph {et~al.}(2005)\citenamefont {Clark},
  \citenamefont {Segall}, \citenamefont {Pickard}, \citenamefont {Hasnip},
  \citenamefont {Probert}, \citenamefont {Refson},\ and\ \citenamefont
  {Payne}}]{CASTEP}%
  \BibitemOpen
  \bibfield  {author} {\bibinfo {author} {\bibfnamefont {Stewart~J.}\
  \bibnamefont {Clark}}, \bibinfo {author} {\bibfnamefont {Matthew~D.}\
  \bibnamefont {Segall}}, \bibinfo {author} {\bibfnamefont {Chris~J.}\
  \bibnamefont {Pickard}}, \bibinfo {author} {\bibfnamefont {Phil~J.}\
  \bibnamefont {Hasnip}}, \bibinfo {author} {\bibfnamefont {Matt I.~J.}\
  \bibnamefont {Probert}}, \bibinfo {author} {\bibfnamefont {Keith}\
  \bibnamefont {Refson}}, \ and\ \bibinfo {author} {\bibfnamefont {Mike~C.}\
  \bibnamefont {Payne}},\ }\bibfield  {title} {\enquote {\bibinfo {title}
  {First principles methods using {\sc castep}},}\ }\href
  {http://www.degruyter.com/view/j/zkri.2005.220.issue-5-6/zkri.220.5.567.65075/zkri.220.5.567.65075.xml}
  {\bibfield  {journal} {\bibinfo  {journal} {Z. Kristallogr.}\ }\textbf
  {\bibinfo {volume} {220}},\ \bibinfo {pages} {567} (\bibinfo {year}
  {2005})}\BibitemShut {NoStop}%
\bibitem [{\citenamefont {Pickard}\ and\ \citenamefont
  {Needs}(2011)}]{Pickard2011}%
  \BibitemOpen
  \bibfield  {author} {\bibinfo {author} {\bibfnamefont {Chris~J.}\
  \bibnamefont {Pickard}}\ and\ \bibinfo {author} {\bibfnamefont {R.~J.}\
  \bibnamefont {Needs}},\ }\bibfield  {title} {\enquote {\bibinfo {title}
  {\textit{Ab initio} random structure searching},}\ }\href
  {http://stacks.iop.org/0953-8984/23/i=5/a=053201} {\bibfield  {journal}
  {\bibinfo  {journal} {J. Phys. Condens. Matter}\ }\textbf {\bibinfo {volume}
  {23}},\ \bibinfo {pages} {053201} (\bibinfo {year} {2011})}\BibitemShut
  {NoStop}%
\bibitem [{\citenamefont {Lloyd-Williams}\ and\ \citenamefont
  {Monserrat}(2015)}]{non_diagonal}%
  \BibitemOpen
  \bibfield  {author} {\bibinfo {author} {\bibfnamefont {Jonathan~H.}\
  \bibnamefont {Lloyd-Williams}}\ and\ \bibinfo {author} {\bibfnamefont
  {Bartomeu}\ \bibnamefont {Monserrat}},\ }\bibfield  {title} {\enquote
  {\bibinfo {title} {Lattice dynamics and electron-phonon coupling calculations
  using nondiagonal supercells},}\ }\href
  {http://link.aps.org/doi/10.1103/PhysRevB.92.184301} {\bibfield  {journal}
  {\bibinfo  {journal} {Phys. Rev. B}\ }\textbf {\bibinfo {volume} {92}},\
  \bibinfo {pages} {184301} (\bibinfo {year} {2015})}\BibitemShut {NoStop}%
\bibitem [{sup()}]{supp_fehe}%
  \BibitemOpen
  \href@noop {} {}\bibinfo {note} {See Supplemental Material at [\ldots] for
  numerial details of the first principles calculations and for cif files for
  the FeHe and FeHe$_2$ structures. It includes
  Refs.~\cite{PhysRevLett.77.3865,PhysRevLett.97.045504,phonon_finite_displacement}.}\BibitemShut
  {Stop}%
\bibitem [{\citenamefont {Stixrude}(2012)}]{stixrude_iron_phase_diagram}%
  \BibitemOpen
  \bibfield  {author} {\bibinfo {author} {\bibfnamefont {Lars}\ \bibnamefont
  {Stixrude}},\ }\bibfield  {title} {\enquote {\bibinfo {title} {Structure of
  iron to 1 {G}bar and 40 000 {K}},}\ }\href
  {http://link.aps.org/doi/10.1103/PhysRevLett.108.055505} {\bibfield
  {journal} {\bibinfo  {journal} {Phys. Rev. Lett.}\ }\textbf {\bibinfo
  {volume} {108}},\ \bibinfo {pages} {055505} (\bibinfo {year}
  {2012})}\BibitemShut {NoStop}%
\bibitem [{\citenamefont {Belonoshko}\ \emph {et~al.}(2006)\citenamefont
  {Belonoshko}, \citenamefont {Skorodumova}, \citenamefont {Rosengren},\ and\
  \citenamefont {Johansson}}]{melting_z_method}%
  \BibitemOpen
  \bibfield  {author} {\bibinfo {author} {\bibfnamefont {A.~B.}\ \bibnamefont
  {Belonoshko}}, \bibinfo {author} {\bibfnamefont {N.~V.}\ \bibnamefont
  {Skorodumova}}, \bibinfo {author} {\bibfnamefont {A.}~\bibnamefont
  {Rosengren}}, \ and\ \bibinfo {author} {\bibfnamefont {B.}~\bibnamefont
  {Johansson}},\ }\bibfield  {title} {\enquote {\bibinfo {title} {Melting and
  critical superheating},}\ }\href {\doibase 10.1103/PhysRevB.73.012201}
  {\bibfield  {journal} {\bibinfo  {journal} {Phys. Rev. B}\ }\textbf {\bibinfo
  {volume} {73}},\ \bibinfo {pages} {012201} (\bibinfo {year}
  {2006})}\BibitemShut {NoStop}%
\bibitem [{\citenamefont {Giannozzi}\ \emph {et~al.}(2009)\citenamefont
  {Giannozzi}, \citenamefont {Baroni}, \citenamefont {Bonini}, \citenamefont
  {Calandra}, \citenamefont {Car}, \citenamefont {Cavazzoni}, \citenamefont
  {Ceresoli}, \citenamefont {Chiarotti}, \citenamefont {Cococcioni},
  \citenamefont {Dabo}, \citenamefont {Corso}, \citenamefont {de~Gironcoli},
  \citenamefont {Fabris}, \citenamefont {Fratesi}, \citenamefont {Gebauer},
  \citenamefont {Gerstmann}, \citenamefont {Gougoussis}, \citenamefont
  {Kokalj}, \citenamefont {Lazzeri}, \citenamefont {Martin-Samos},
  \citenamefont {Marzari}, \citenamefont {Mauri}, \citenamefont {Mazzarello},
  \citenamefont {Paolini}, \citenamefont {Pasquarello}, \citenamefont
  {Paulatto}, \citenamefont {Sbraccia}, \citenamefont {Scandolo}, \citenamefont
  {Sclauzero}, \citenamefont {Seitsonen}, \citenamefont {Smogunov},
  \citenamefont {Umari},\ and\ \citenamefont
  {Wentzcovitch}}]{quantum_espresso}%
  \BibitemOpen
  \bibfield  {author} {\bibinfo {author} {\bibfnamefont {Paolo}\ \bibnamefont
  {Giannozzi}}, \bibinfo {author} {\bibfnamefont {Stefano}\ \bibnamefont
  {Baroni}}, \bibinfo {author} {\bibfnamefont {Nicola}\ \bibnamefont {Bonini}},
  \bibinfo {author} {\bibfnamefont {Matteo}\ \bibnamefont {Calandra}}, \bibinfo
  {author} {\bibfnamefont {Roberto}\ \bibnamefont {Car}}, \bibinfo {author}
  {\bibfnamefont {Carlo}\ \bibnamefont {Cavazzoni}}, \bibinfo {author}
  {\bibfnamefont {Davide}\ \bibnamefont {Ceresoli}}, \bibinfo {author}
  {\bibfnamefont {Guido~L}\ \bibnamefont {Chiarotti}}, \bibinfo {author}
  {\bibfnamefont {Matteo}\ \bibnamefont {Cococcioni}}, \bibinfo {author}
  {\bibfnamefont {Ismaila}\ \bibnamefont {Dabo}}, \bibinfo {author}
  {\bibfnamefont {Andrea~Dal}\ \bibnamefont {Corso}}, \bibinfo {author}
  {\bibfnamefont {Stefano}\ \bibnamefont {de~Gironcoli}}, \bibinfo {author}
  {\bibfnamefont {Stefano}\ \bibnamefont {Fabris}}, \bibinfo {author}
  {\bibfnamefont {Guido}\ \bibnamefont {Fratesi}}, \bibinfo {author}
  {\bibfnamefont {Ralph}\ \bibnamefont {Gebauer}}, \bibinfo {author}
  {\bibfnamefont {Uwe}\ \bibnamefont {Gerstmann}}, \bibinfo {author}
  {\bibfnamefont {Christos}\ \bibnamefont {Gougoussis}}, \bibinfo {author}
  {\bibfnamefont {Anton}\ \bibnamefont {Kokalj}}, \bibinfo {author}
  {\bibfnamefont {Michele}\ \bibnamefont {Lazzeri}}, \bibinfo {author}
  {\bibfnamefont {Layla}\ \bibnamefont {Martin-Samos}}, \bibinfo {author}
  {\bibfnamefont {Nicola}\ \bibnamefont {Marzari}}, \bibinfo {author}
  {\bibfnamefont {Francesco}\ \bibnamefont {Mauri}}, \bibinfo {author}
  {\bibfnamefont {Riccardo}\ \bibnamefont {Mazzarello}}, \bibinfo {author}
  {\bibfnamefont {Stefano}\ \bibnamefont {Paolini}}, \bibinfo {author}
  {\bibfnamefont {Alfredo}\ \bibnamefont {Pasquarello}}, \bibinfo {author}
  {\bibfnamefont {Lorenzo}\ \bibnamefont {Paulatto}}, \bibinfo {author}
  {\bibfnamefont {Carlo}\ \bibnamefont {Sbraccia}}, \bibinfo {author}
  {\bibfnamefont {Sandro}\ \bibnamefont {Scandolo}}, \bibinfo {author}
  {\bibfnamefont {Gabriele}\ \bibnamefont {Sclauzero}}, \bibinfo {author}
  {\bibfnamefont {Ari~P}\ \bibnamefont {Seitsonen}}, \bibinfo {author}
  {\bibfnamefont {Alexander}\ \bibnamefont {Smogunov}}, \bibinfo {author}
  {\bibfnamefont {Paolo}\ \bibnamefont {Umari}}, \ and\ \bibinfo {author}
  {\bibfnamefont {Renata~M}\ \bibnamefont {Wentzcovitch}},\ }\bibfield  {title}
  {\enquote {\bibinfo {title} {{QUANTUM ESPRESSO}: a modular and open-source
  software project for quantum simulations of materials},}\ }\href
  {http://stacks.iop.org/0953-8984/21/i=39/a=395502} {\bibfield  {journal}
  {\bibinfo  {journal} {J. Phys.: Condens. Matter}\ }\textbf {\bibinfo {volume}
  {21}},\ \bibinfo {pages} {395502} (\bibinfo {year} {2009})}\BibitemShut
  {NoStop}%
\bibitem [{\citenamefont {Hull}(2004)}]{superionic_review}%
  \BibitemOpen
  \bibfield  {author} {\bibinfo {author} {\bibfnamefont {Stephen}\ \bibnamefont
  {Hull}},\ }\bibfield  {title} {\enquote {\bibinfo {title} {Superionics:
  crystal structures and conduction processes},}\ }\href
  {http://stacks.iop.org/0034-4885/67/i=7/a=R05} {\bibfield  {journal}
  {\bibinfo  {journal} {Rep. Prog. Phys.}\ }\textbf {\bibinfo {volume} {67}},\
  \bibinfo {pages} {1233} (\bibinfo {year} {2004})}\BibitemShut {NoStop}%
\bibitem [{\citenamefont {Kamaya}\ \emph {et~al.}(2011)\citenamefont {Kamaya},
  \citenamefont {Homma}, \citenamefont {Yamakawa}, \citenamefont {Hirayama},
  \citenamefont {Kanno}, \citenamefont {Yonemura}, \citenamefont {Kamiyama},
  \citenamefont {Kato}, \citenamefont {Hama}, \citenamefont {Kawamoto},\ and\
  \citenamefont {Mitsui}}]{lithium_superionic_conductor}%
  \BibitemOpen
  \bibfield  {author} {\bibinfo {author} {\bibfnamefont {Noriaki}\ \bibnamefont
  {Kamaya}}, \bibinfo {author} {\bibfnamefont {Kenji}\ \bibnamefont {Homma}},
  \bibinfo {author} {\bibfnamefont {Yuichiro}\ \bibnamefont {Yamakawa}},
  \bibinfo {author} {\bibfnamefont {Masaaki}\ \bibnamefont {Hirayama}},
  \bibinfo {author} {\bibfnamefont {Ryoji}\ \bibnamefont {Kanno}}, \bibinfo
  {author} {\bibfnamefont {Masao}\ \bibnamefont {Yonemura}}, \bibinfo {author}
  {\bibfnamefont {Takashi}\ \bibnamefont {Kamiyama}}, \bibinfo {author}
  {\bibfnamefont {Yuki}\ \bibnamefont {Kato}}, \bibinfo {author} {\bibfnamefont
  {Shigenori}\ \bibnamefont {Hama}}, \bibinfo {author} {\bibfnamefont {Koji}\
  \bibnamefont {Kawamoto}}, \ and\ \bibinfo {author} {\bibfnamefont {Akio}\
  \bibnamefont {Mitsui}},\ }\bibfield  {title} {\enquote {\bibinfo {title} {A
  lithium superionic conductor},}\ }\href {http://dx.doi.org/10.1038/nmat3066}
  {\bibfield  {journal} {\bibinfo  {journal} {Nat. Mater.}\ }\textbf {\bibinfo
  {volume} {10}},\ \bibinfo {pages} {682} (\bibinfo {year} {2011})}\BibitemShut
  {NoStop}%
\bibitem [{\citenamefont {French}\ \emph {et~al.}(2010)\citenamefont {French},
  \citenamefont {Mattsson},\ and\ \citenamefont
  {Redmer}}]{superionic_metallic_water}%
  \BibitemOpen
  \bibfield  {author} {\bibinfo {author} {\bibfnamefont {Martin}\ \bibnamefont
  {French}}, \bibinfo {author} {\bibfnamefont {Thomas~R.}\ \bibnamefont
  {Mattsson}}, \ and\ \bibinfo {author} {\bibfnamefont {Ronald}\ \bibnamefont
  {Redmer}},\ }\bibfield  {title} {\enquote {\bibinfo {title} {Diffusion and
  electrical conductivity in water at ultrahigh pressures},}\ }\href
  {https://link.aps.org/doi/10.1103/PhysRevB.82.174108} {\bibfield  {journal}
  {\bibinfo  {journal} {Phys. Rev. B}\ }\textbf {\bibinfo {volume} {82}},\
  \bibinfo {pages} {174108} (\bibinfo {year} {2010})}\BibitemShut {NoStop}%
\bibitem [{\citenamefont {Morard}\ \emph {et~al.}(2011)\citenamefont {Morard},
  \citenamefont {Bouchet}, \citenamefont {Valencia}, \citenamefont {Mazevet},\
  and\ \citenamefont {Guyot}}]{iron_melting_dft_tpa}%
  \BibitemOpen
  \bibfield  {author} {\bibinfo {author} {\bibfnamefont {G.}~\bibnamefont
  {Morard}}, \bibinfo {author} {\bibfnamefont {J.}~\bibnamefont {Bouchet}},
  \bibinfo {author} {\bibfnamefont {D.}~\bibnamefont {Valencia}}, \bibinfo
  {author} {\bibfnamefont {S.}~\bibnamefont {Mazevet}}, \ and\ \bibinfo
  {author} {\bibfnamefont {F.}~\bibnamefont {Guyot}},\ }\bibfield  {title}
  {\enquote {\bibinfo {title} {The melting curve of iron at extreme pressures:
  Implications for planetary cores},}\ }\href
  {http://www.sciencedirect.com/science/article/pii/S1574181811000231}
  {\bibfield  {journal} {\bibinfo  {journal} {High Energy Density Phys.}\
  }\textbf {\bibinfo {volume} {7}},\ \bibinfo {pages} {141--144} (\bibinfo
  {year} {2011})}\BibitemShut {NoStop}%
\bibitem [{\citenamefont {Nettelmann}\ \emph {et~al.}(2012)\citenamefont
  {Nettelmann}, \citenamefont {Becker}, \citenamefont {Holst},\ and\
  \citenamefont {Redmer}}]{jupiter_model_nettelmann}%
  \BibitemOpen
  \bibfield  {author} {\bibinfo {author} {\bibfnamefont {N.}~\bibnamefont
  {Nettelmann}}, \bibinfo {author} {\bibfnamefont {A.}~\bibnamefont {Becker}},
  \bibinfo {author} {\bibfnamefont {B.}~\bibnamefont {Holst}}, \ and\ \bibinfo
  {author} {\bibfnamefont {R.}~\bibnamefont {Redmer}},\ }\bibfield  {title}
  {\enquote {\bibinfo {title} {Jupiter models with improved ab initio hydrogen
  equation of state ({H-REOS.2})},}\ }\href
  {http://stacks.iop.org/0004-637X/750/i=1/a=52} {\bibfield  {journal}
  {\bibinfo  {journal} {Astrophys. J.}\ }\textbf {\bibinfo {volume} {750}},\
  \bibinfo {pages} {52} (\bibinfo {year} {2012})}\BibitemShut {NoStop}%
\bibitem [{\citenamefont {Stevenson}(1980)}]{saturn_atmosphere}%
  \BibitemOpen
  \bibfield  {author} {\bibinfo {author} {\bibfnamefont {D.~J.}\ \bibnamefont
  {Stevenson}},\ }\bibfield  {title} {\enquote {\bibinfo {title} {Saturn's
  luminosity and magnetism},}\ }\href
  {http://science.sciencemag.org/content/208/4445/746} {\bibfield  {journal}
  {\bibinfo  {journal} {Science}\ }\textbf {\bibinfo {volume} {208}},\ \bibinfo
  {pages} {746--748} (\bibinfo {year} {1980})}\BibitemShut {NoStop}%
\bibitem [{\citenamefont {Hansen}\ and\ \citenamefont
  {Liebert}(2003)}]{cool_wds}%
  \BibitemOpen
  \bibfield  {author} {\bibinfo {author} {\bibfnamefont {Brad~M.S.}\
  \bibnamefont {Hansen}}\ and\ \bibinfo {author} {\bibfnamefont {James}\
  \bibnamefont {Liebert}},\ }\bibfield  {title} {\enquote {\bibinfo {title}
  {Cool white dwarfs},}\ }\href
  {https://doi.org/10.1146/annurev.astro.41.081401.155117} {\bibfield
  {journal} {\bibinfo  {journal} {Annu. Rev. Astron. Astrophys.}\ }\textbf
  {\bibinfo {volume} {41}},\ \bibinfo {pages} {465--515} (\bibinfo {year}
  {2003})}\BibitemShut {NoStop}%
\bibitem [{\citenamefont {Perdew}\ \emph {et~al.}(1996)\citenamefont {Perdew},
  \citenamefont {Burke},\ and\ \citenamefont
  {Ernzerhof}}]{PhysRevLett.77.3865}%
  \BibitemOpen
  \bibfield  {author} {\bibinfo {author} {\bibfnamefont {John~P.}\ \bibnamefont
  {Perdew}}, \bibinfo {author} {\bibfnamefont {Kieron}\ \bibnamefont {Burke}},
  \ and\ \bibinfo {author} {\bibfnamefont {Matthias}\ \bibnamefont
  {Ernzerhof}},\ }\bibfield  {title} {\enquote {\bibinfo {title} {Generalized
  gradient approximation made simple},}\ }\href
  {http://link.aps.org/doi/10.1103/PhysRevLett.77.3865} {\bibfield  {journal}
  {\bibinfo  {journal} {Phys. Rev. Lett.}\ }\textbf {\bibinfo {volume} {77}},\
  \bibinfo {pages} {3865} (\bibinfo {year} {1996})}\BibitemShut {NoStop}%
\bibitem [{\citenamefont {Pickard}\ and\ \citenamefont
  {Needs}(2006)}]{PhysRevLett.97.045504}%
  \BibitemOpen
  \bibfield  {author} {\bibinfo {author} {\bibfnamefont {Chris~J.}\
  \bibnamefont {Pickard}}\ and\ \bibinfo {author} {\bibfnamefont {R.~J.}\
  \bibnamefont {Needs}},\ }\bibfield  {title} {\enquote {\bibinfo {title}
  {High-pressure phases of silane},}\ }\href
  {http://link.aps.org/doi/10.1103/PhysRevLett.97.045504} {\bibfield  {journal}
  {\bibinfo  {journal} {Phys. Rev. Lett.}\ }\textbf {\bibinfo {volume} {97}},\
  \bibinfo {pages} {045504} (\bibinfo {year} {2006})}\BibitemShut {NoStop}%
\bibitem [{\citenamefont {Kunc}\ and\ \citenamefont
  {Martin}(1982)}]{phonon_finite_displacement}%
  \BibitemOpen
  \bibfield  {author} {\bibinfo {author} {\bibfnamefont {K.}~\bibnamefont
  {Kunc}}\ and\ \bibinfo {author} {\bibfnamefont {Richard~M.}\ \bibnamefont
  {Martin}},\ }\bibfield  {title} {\enquote {\bibinfo {title} {\textit{Ab
  Initio} force constants of {G}a{A}s: A new approach to calculation of phonons
  and dielectric properties},}\ }\href
  {http://link.aps.org/doi/10.1103/PhysRevLett.48.406} {\bibfield  {journal}
  {\bibinfo  {journal} {Phys. Rev. Lett.}\ }\textbf {\bibinfo {volume} {48}},\
  \bibinfo {pages} {406--409} (\bibinfo {year} {1982})}\BibitemShut {NoStop}%
\end{thebibliography}%

\onecolumngrid
\clearpage
\begin{center}
\textbf{\large Supplemental Material for ``Helium-iron compounds at terapascal pressures''}
\end{center}
\setcounter{equation}{0}
\setcounter{figure}{0}
\setcounter{table}{0}
\setcounter{page}{1}
\makeatletter
\renewcommand{\theequation}{S\arabic{equation}}
\renewcommand{\thefigure}{S\arabic{figure}}
\renewcommand{\bibnumfmt}[1]{[S#1]}
\renewcommand{\citenumfont}[1]{S#1}
\section{Computational details}

\subsection{Density functional theory}
Density functional theory calculations have been performed using the plane-wave pseudopotential {\sc castep} code~\cite{CASTEP} with the PBE functional~\cite{PhysRevLett.77.3865}. The extreme pressures we study force us to use tailored pseudopotentials with core radii of $0.6$~bohr for helium and $1.0$~bohr for iron, and our calculations explicitly include the $1s$ states of helium and the $3s3p3d4s$ states of iron. The pseudopotential strings used are:
\begin{verbatim}
He 1|0.6|27|31|36|10(qc=11)
Fe 3|1.0|33|38|44|30U:40:31:32(qc=11)
\end{verbatim}
These hard pseudopotentials require plane-wave cut-off energies of $1500$~eV. For the searches we have used Brillouin zone $\mathbf{k}$-point grids of density $2\pi\times0.07$~\AA$^{-1}$, and finer grids of density $2\pi\times0.025$~\AA$^{-1}$ for the final results and the vibrational calculations. Structures were relaxed to achieve uncertainties in the pressure below $0.1$~TPa.

\subsection{AIRSS}
The structure searches have been performed using the AIRSS method~\cite{PhysRevLett.97.045504,Pickard2011} at pressures of $5$~TPa, $10$~TPa, and $50$~TPa, and with stoichiometries Fe$_x$He$_y$ for $x,y=1,\ldots,6$. The total number of structures generated in the searches is $7,371$.

\subsection{Vibrational calculations}

Harmonic vibrational free energies have been calculated using finite-displacements~\cite{phonon_finite_displacement} with the nondiagonal supercells method~\cite{non_diagonal}. The system sizes required to obtain converged results are shown in Table~\ref{tab:nondiagonal}, and the accuracy required could not have been achieved unless nondiagonal supercells had been used, as the computational cost of using small radii pseudopotentials is significant. Results accurate to $1$~meV/atom have been obtained by diagonalising the corresponding dynamical matrices over a fine vibrational Brillouin-zone grid. Anharmonic vibrations have been previously found to be negligible in helium at TPa pressures~\cite{Monserrat_2014_he_elph}.

\begin{table}[h]
  \setlength{\tabcolsep}{8pt} 
  \caption{Size of largest supercell required for the corresponding vibrational BZ grid.}
  \label{tab:nondiagonal}
  \begin{ruledtabular}
  \begin{tabular}{c|ccc}
  & BZ grid size & Diagonal supercells & Nondiagonal supercells \\
  \hline
  He-hpc   &  $4\times4\times4$  & $64$ cells ($128$ atoms) & $4$ cells ($8$ atoms) \\
  Fe-hcp   &  $4\times4\times4$  & $64$ cells ($128$ atoms) & $4$ cells ($8$ atoms) \\
  Fe-fcc   &  $4\times4\times4$  & $64$ cells ($64$ atoms) & $4$ cells ($4$ atoms) \\
  FeHe-$Cmcm$     &  $3\times3\times3$  & $27$ cells ($216$ atoms) & $3$ cells ($24$ atoms) \\
  FeHe$_2$-$P6/mmm$ &  $5\times5\times5$  & $125$ cells ($375$ atoms) & $5$ cells ($15$ atoms) \\
  FeHe$_2$-$I4_1/amd$ &  $3\times3\times3$  & $27$ cells ($162$ atoms) & $3$ cells ($18$ atoms) \\
  FeHe$_2$-$Cmmm$ &  $4\times4\times4$  & $64$ cells ($576$ atoms) & $4$ cells ($36$ atoms) \\
  \end{tabular}
  \end{ruledtabular}
\end{table}

\section{Thermodynamic stability}

In this section we show the relative stability of the various crystal structures for iron, helium, FeHe, and FeHe$_2$. For any pressure-temperature conditions, the thermodynamically stable structure is used in the main text.

\begin{figure}
\subfloat[][Static lattice enthalpies.]{
  \includegraphics[scale=0.36]{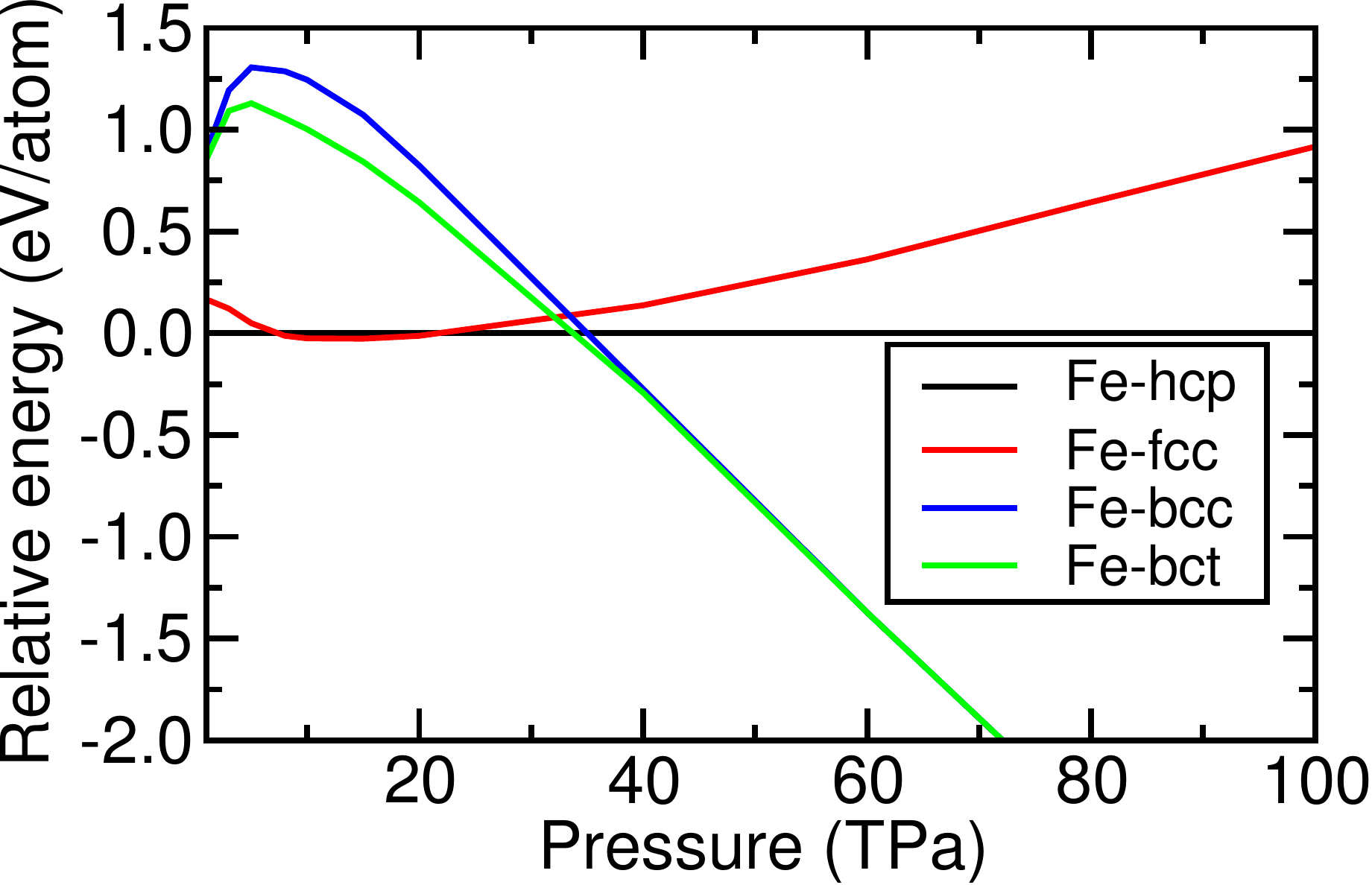}
\label{subfig:fe_static}}
\subfloat[][Gibbs free energies.]{
  \includegraphics[scale=0.36]{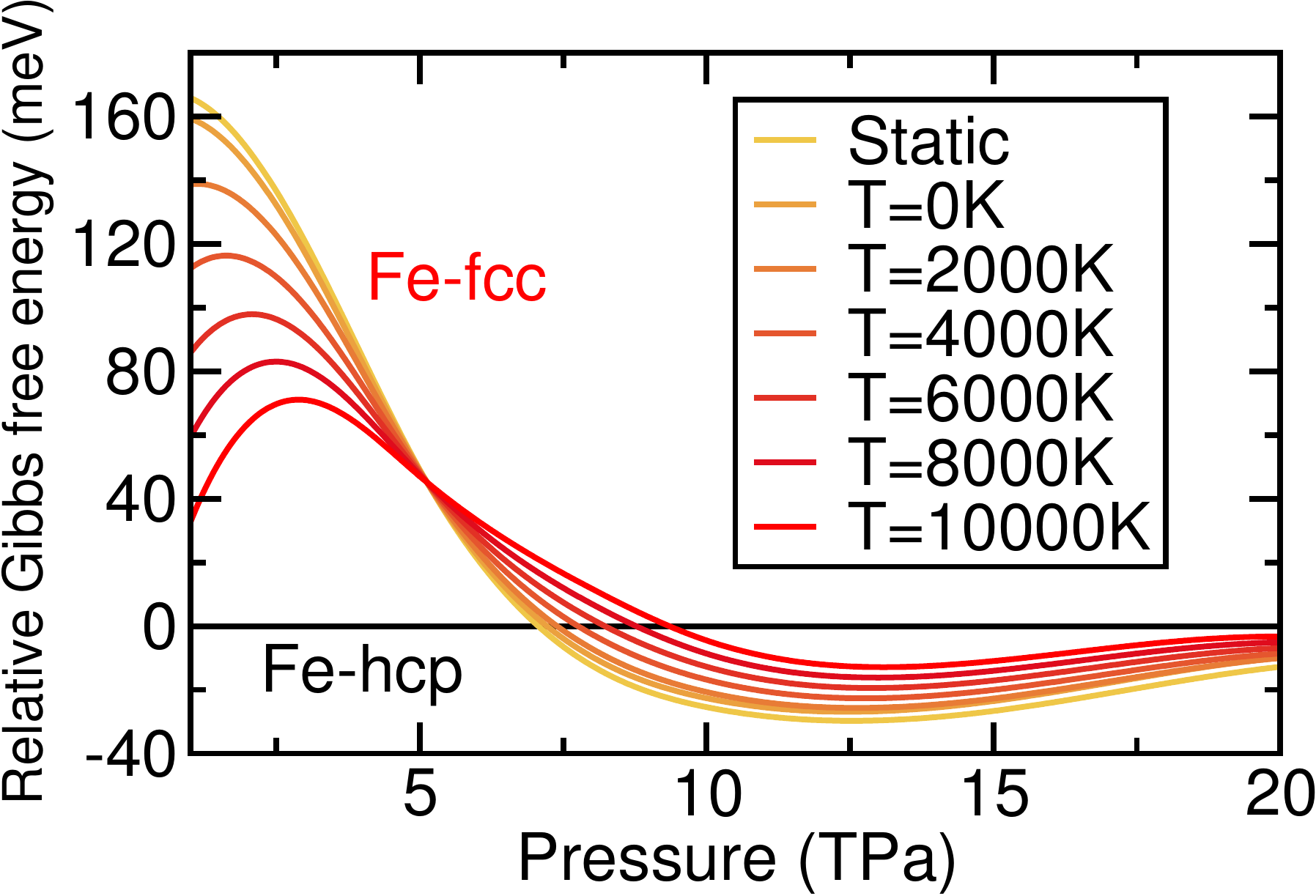}
\label{subfig:fe_gibbs}}
  \caption{Relative enthalpies and Gibbs free energies of fcc and bcc iron with respect to hcp iron.}
      \label{fig:iron}
\end{figure}

The relevant structures of iron are the face-centered cubic (fcc), body-centered cubic (bcc), body-centered tetragonal (bct), and hexagonal closed-packed (hcp). Their relative static lattice enthalpies are shown in Fig.~\ref{subfig:fe_static}. In agreement with previous reports~\cite{Pickard_2009_fe_terapascal}, we find that fcc iron becomes stable compared to hcp iron at pressures in the range $7.4$~TPa to $21.7$~TPa, and a body-centered tetragonal (bct) iron is the most stable structure above $35$~TPa. The bct structure approaches the bcc structure with increasing pressure. The effects of temperature are shown in Fig.~\ref{subfig:fe_gibbs} up to $20$~TPa. Including quantum and thermal nuclear motion destabilises fcc iron, so that at $10,000$~K it only becomes stable at $9.4$~TPa.

\begin{figure}
  \begin{minipage}[b]{0.45\linewidth}
    \centering
      \includegraphics[scale=0.36]{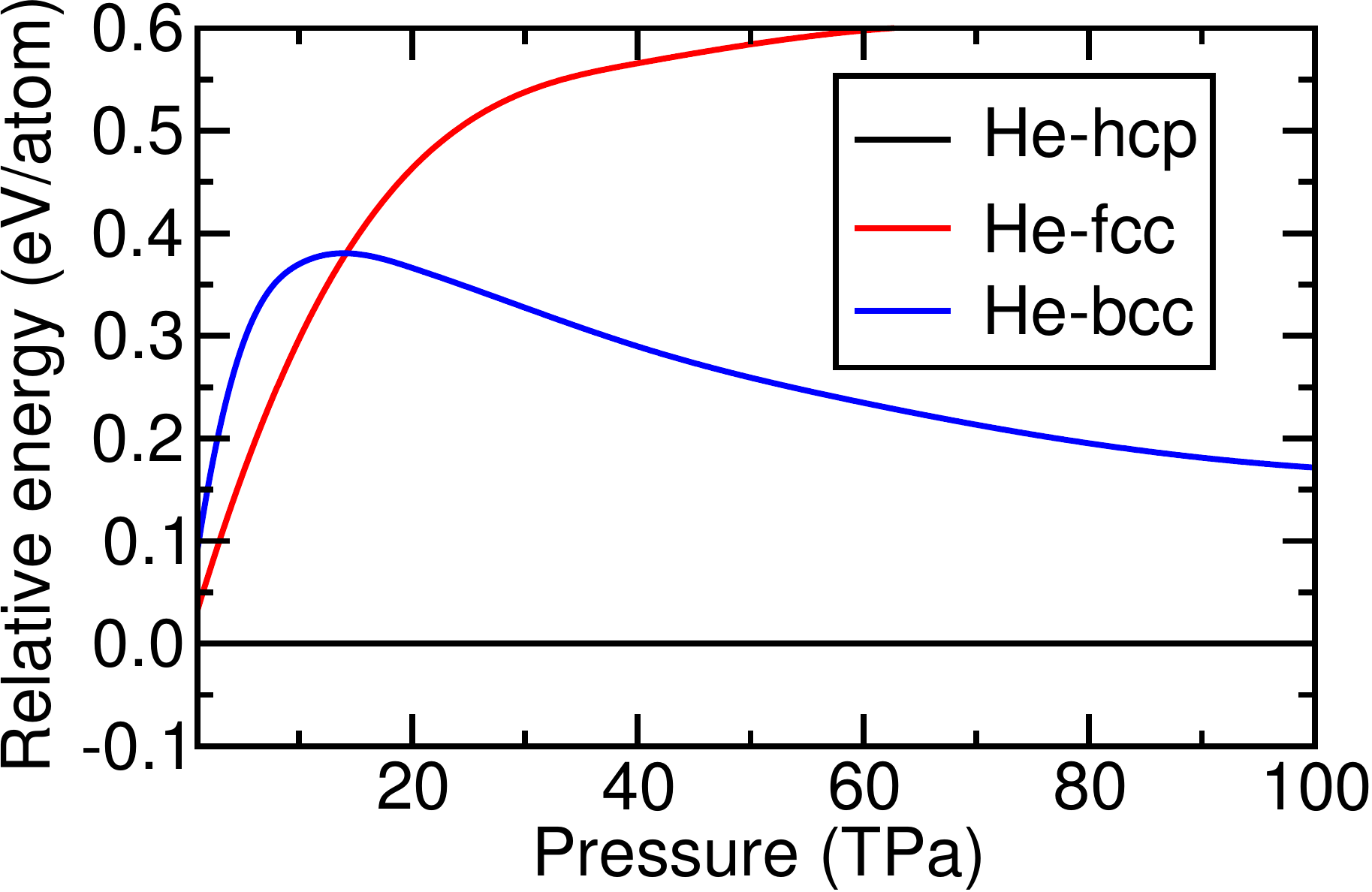}
        \caption{Relative enthalpies of fcc and bcc helium with respect to hcp helium.}
        \label{fig:static_he}
   \end{minipage}
   \hspace{0.5cm}
  \begin{minipage}[b]{0.45\linewidth}
    \centering
     \includegraphics[scale=0.36]{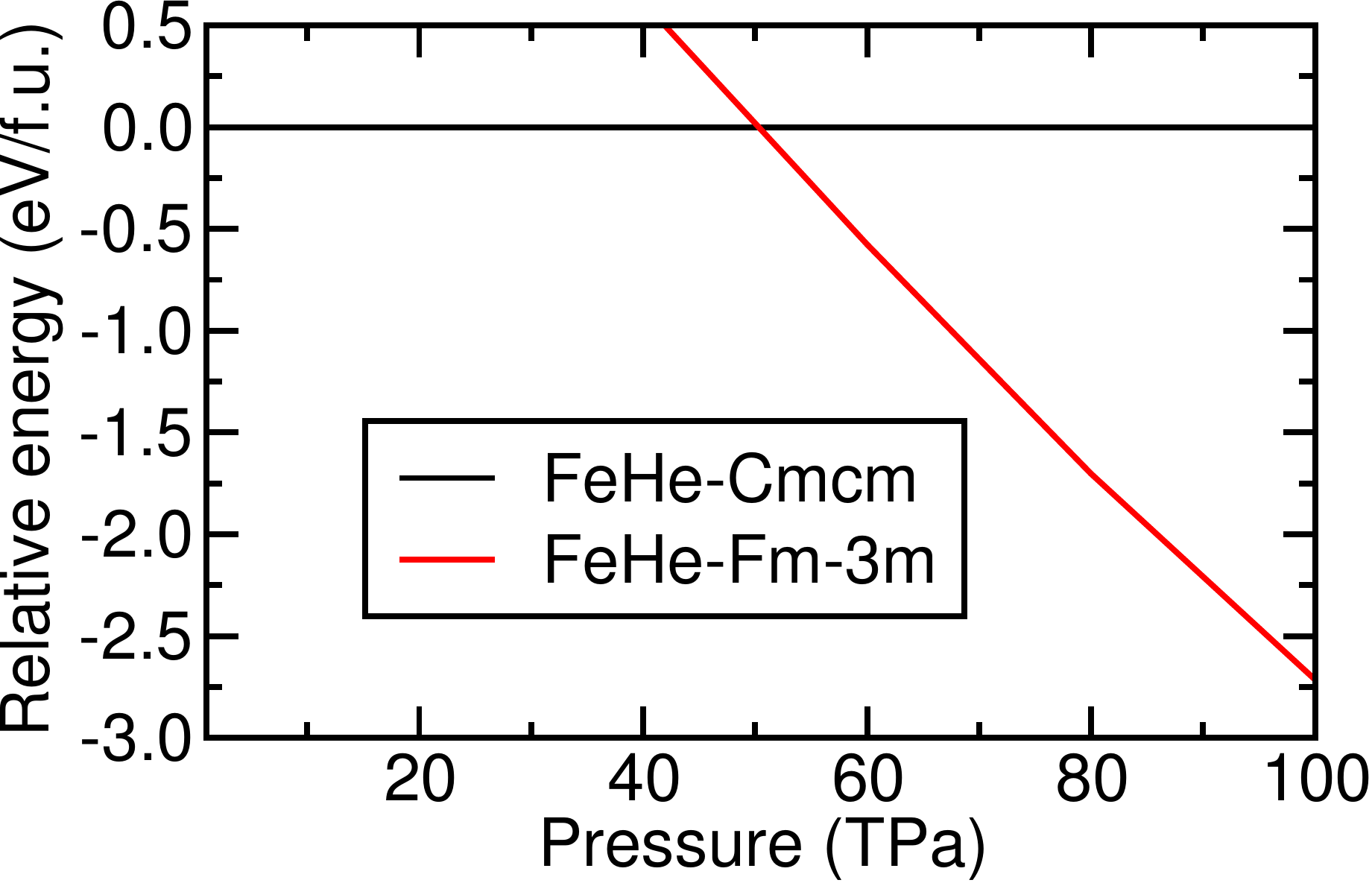}
     \caption{Relative enthalpies of FeHe-$Cmcm$ and FeHe-$Fm$-$3m$.}
      \label{fig:static_fehe}
  \end{minipage}
\end{figure}

\begin{figure}
\subfloat[][FeHe$_2$-$P6/mmm$.]{
  \includegraphics[scale=0.25]{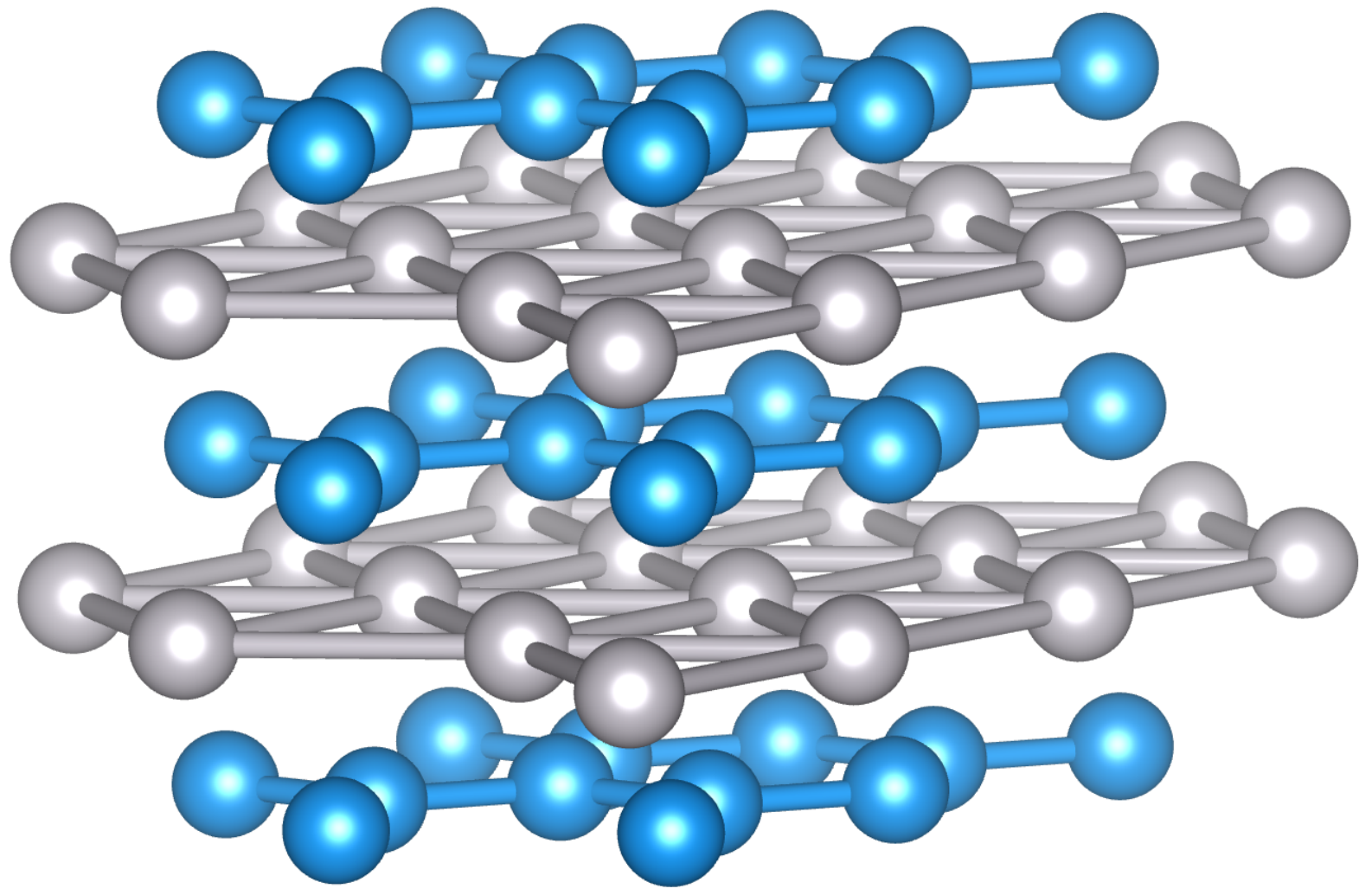}
\label{subfig:fehe}}
\subfloat[][FeHe$_2$-$I4_1/amd$.]{
  \includegraphics[scale=0.25]{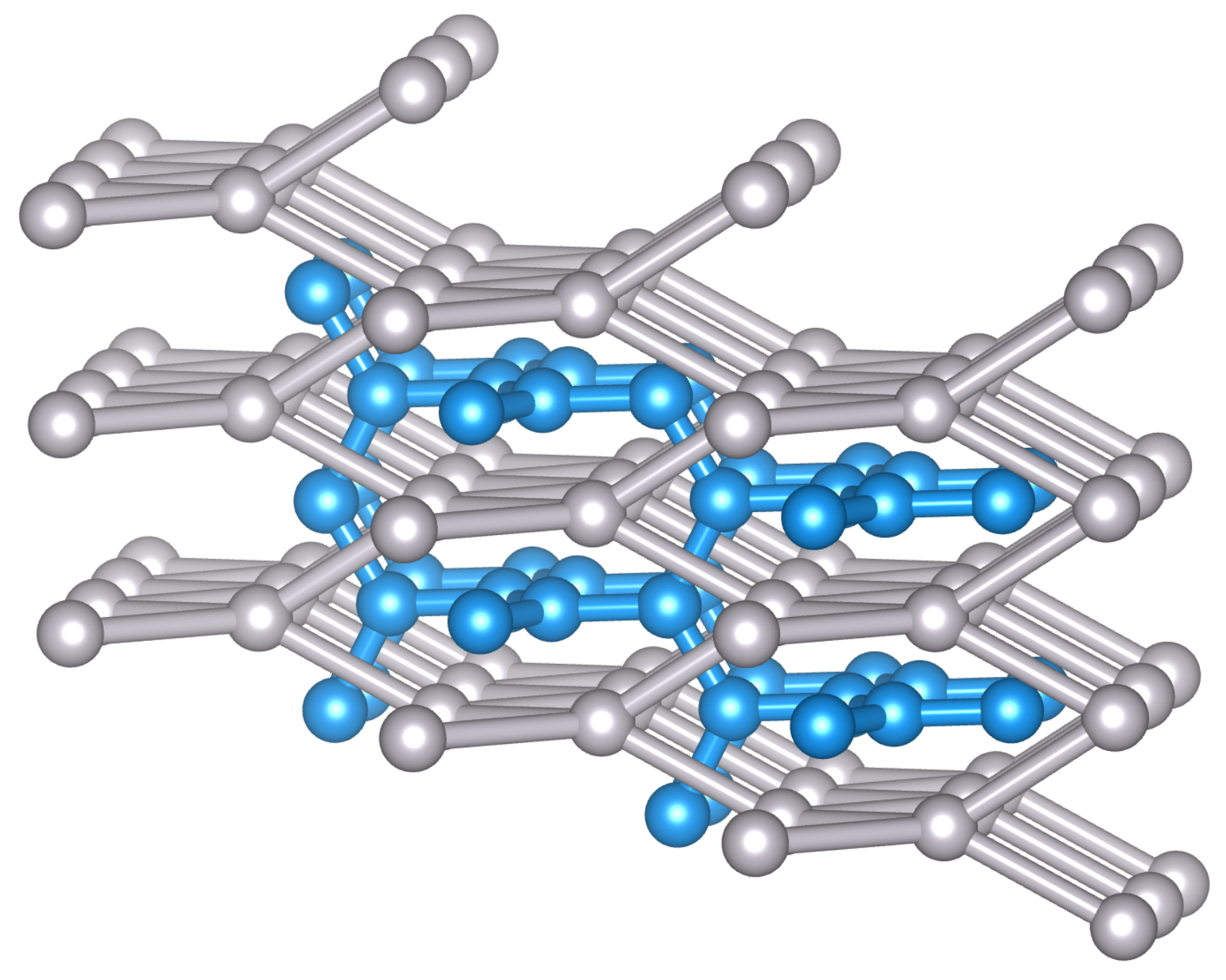}
\label{subfig:fehe2}}
\subfloat[][FeHe$_2$-$Cmmm$.]{
  \includegraphics[scale=0.3]{fehe2-9.pdf}
\label{subfig:fehe2}}
  \caption{Crystal structures of $P6/mmm$ FeHe$_2$, $I4_1/amd$ FeHe$_2$, and $Cmmm$ FeHe$_2$ at $10$~TPa. Helium atoms are represented in blue, and iron atoms in grey.}
      \label{fig:structures}
\end{figure}

\begin{figure}
\subfloat[][Relative enthalpies.]{
  \includegraphics[scale=0.36]{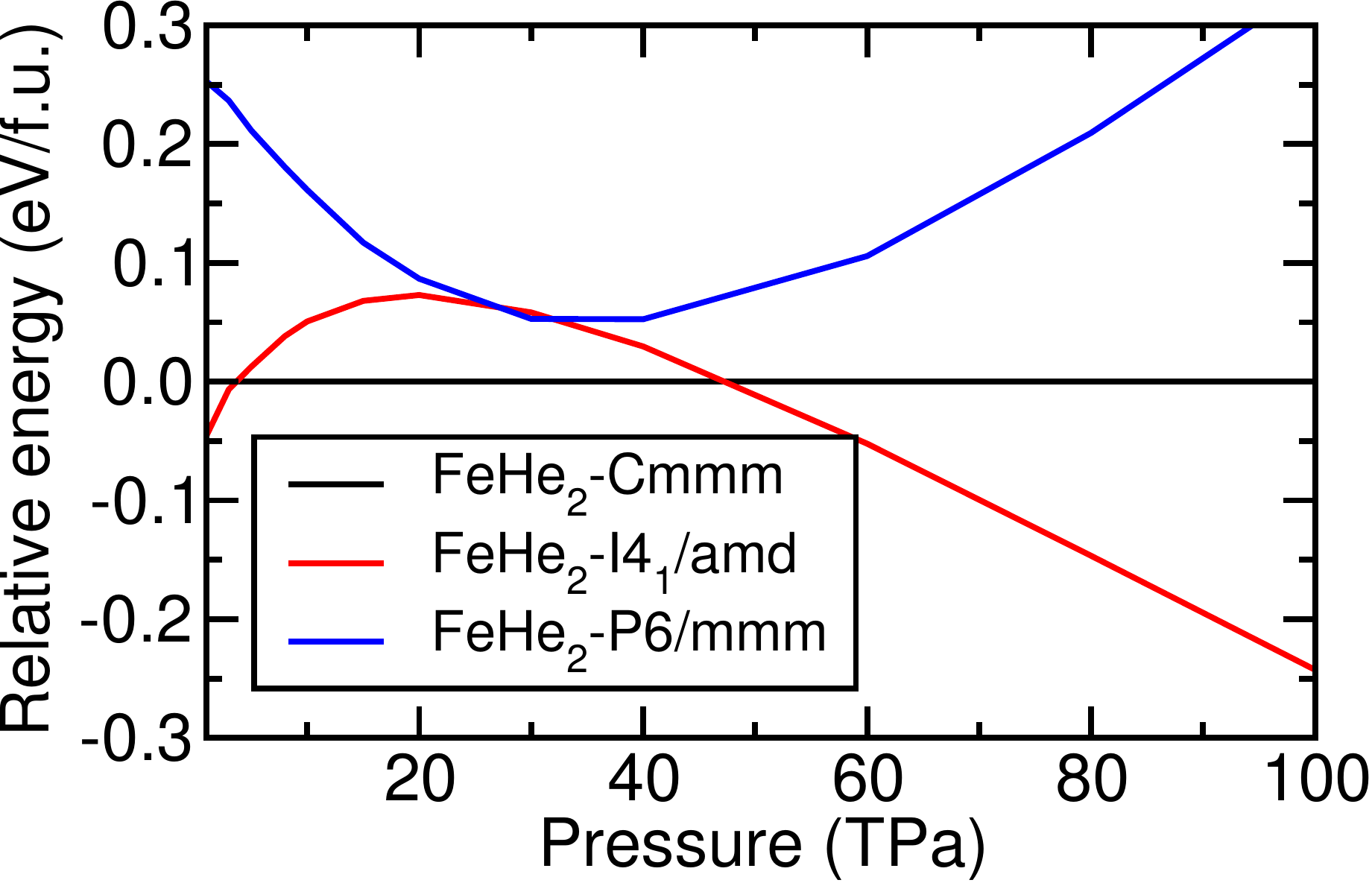}
\label{subfig:static_fehe2}}
\subfloat[][Pressure-temperature phase diagram.]{
  \includegraphics[scale=0.36]{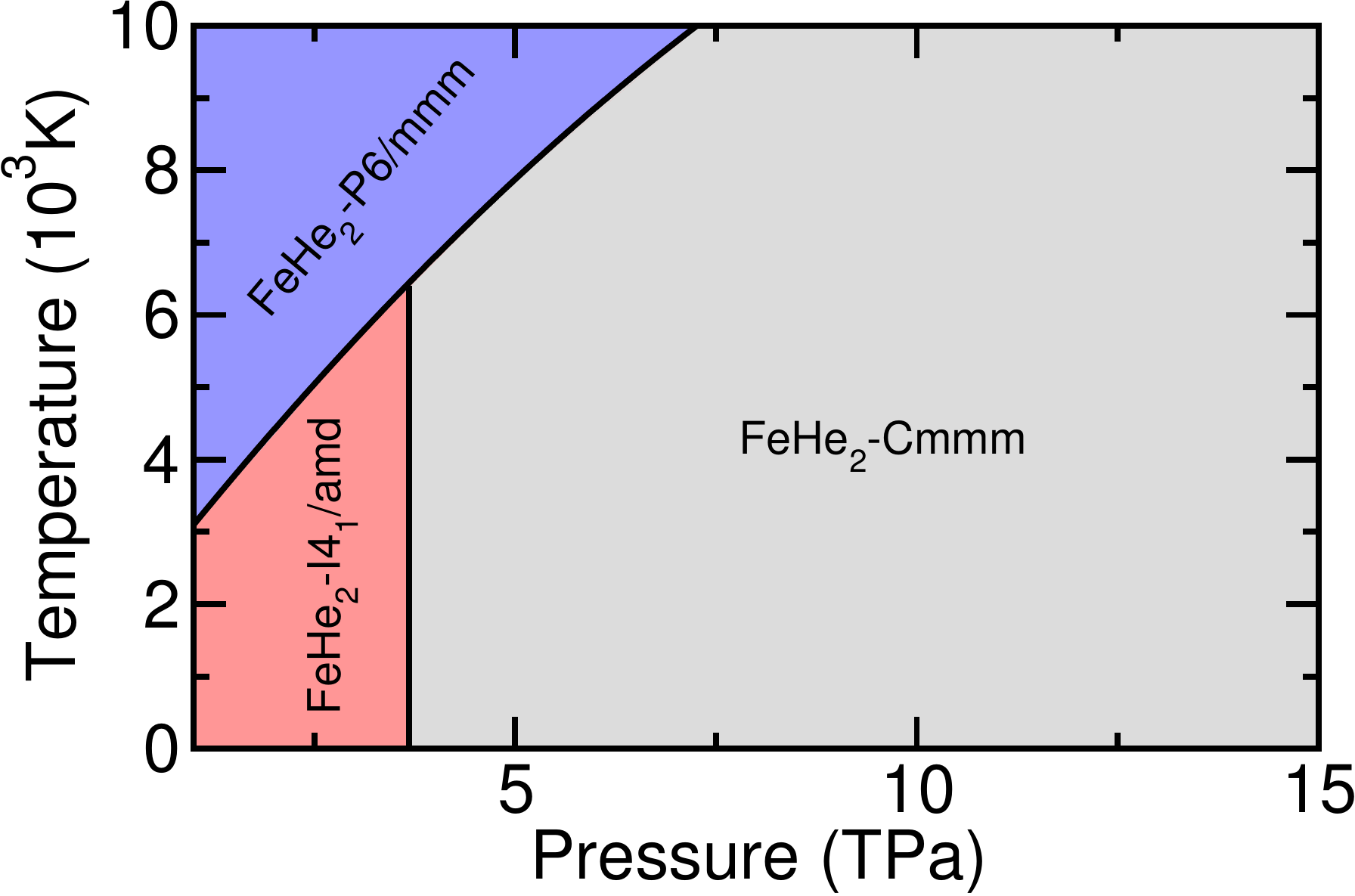}
\label{subfig:fehe2_phase_diagram}}
  \caption{Phase diagram for the helium-iron compounds of FeHe$_2$ stoichiometry.}
      \label{fig:fehe2}
\end{figure}

\begin{figure}
  \includegraphics[scale=0.36]{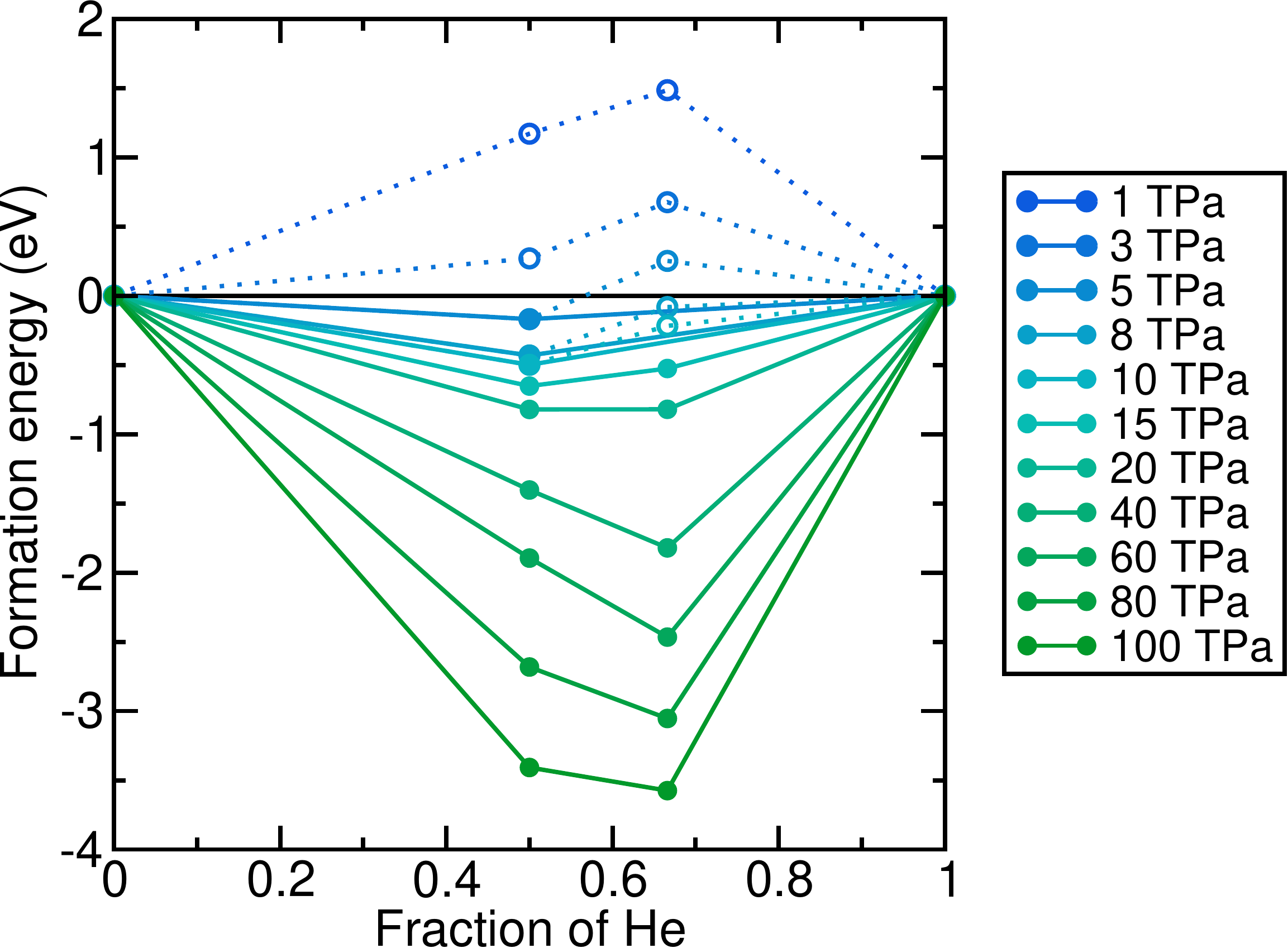}
  \caption{Convex hull diagram for the helium-iron system at the static lattice level and for a range of temperatures. The solid circles indicate structures on the convex hull, and the solid lines connect them. The open circles indicate structures that are not on the convex hull, and these are connected by dotted lines as a guide to the eye.}
      \label{fig:hull}
\end{figure}

Helium is only found in the hcp structure in the entire pressure range from $1$~TPa to $100$~TPa. A comparison with the bcc and fcc structures is provided in Fig.~\ref{fig:static_he}.

For the helium-iron FeHe stoichiometry, there are two different structures that are energetically competitive: (i) a structure of space group $Cmcm$ with $8$ atoms in the primitive cell, and (ii) a structure of space group $Fm\overline{3}m$ with $2$ atoms in the primitive cell (rock-salt structure). The relative enthalpies of these structures are shown in Fig.~\ref{fig:static_fehe}.

For the helium-iron FeHe$_2$ stoichiometry, there are three different structures that are energetically competitive: (i) a structure of space group $P6/mmm$ with $3$ atoms in the primitive cell, (ii) a structure of space group $I4_1/amd$ with $6$ atoms in the primitive cell, and (iii) a structure of space group $Cmmm$ with $9$ atoms in the primitive cell. These structures are depicted in Fig.~\ref{fig:structures}, and their relative enthalpies and low pressure phase diagram are shown in Fig.~\ref{fig:fehe2}.
Combining the He, Fe, FeHe, and FeHe$_2$ results, we construct the convex hull diagram for the helium-iron mixtures, shown in Fig.~\ref{fig:hull}.

\section{Superionicity and melting}

\begin{figure}[h!]
\centering
  \includegraphics[scale=0.30]{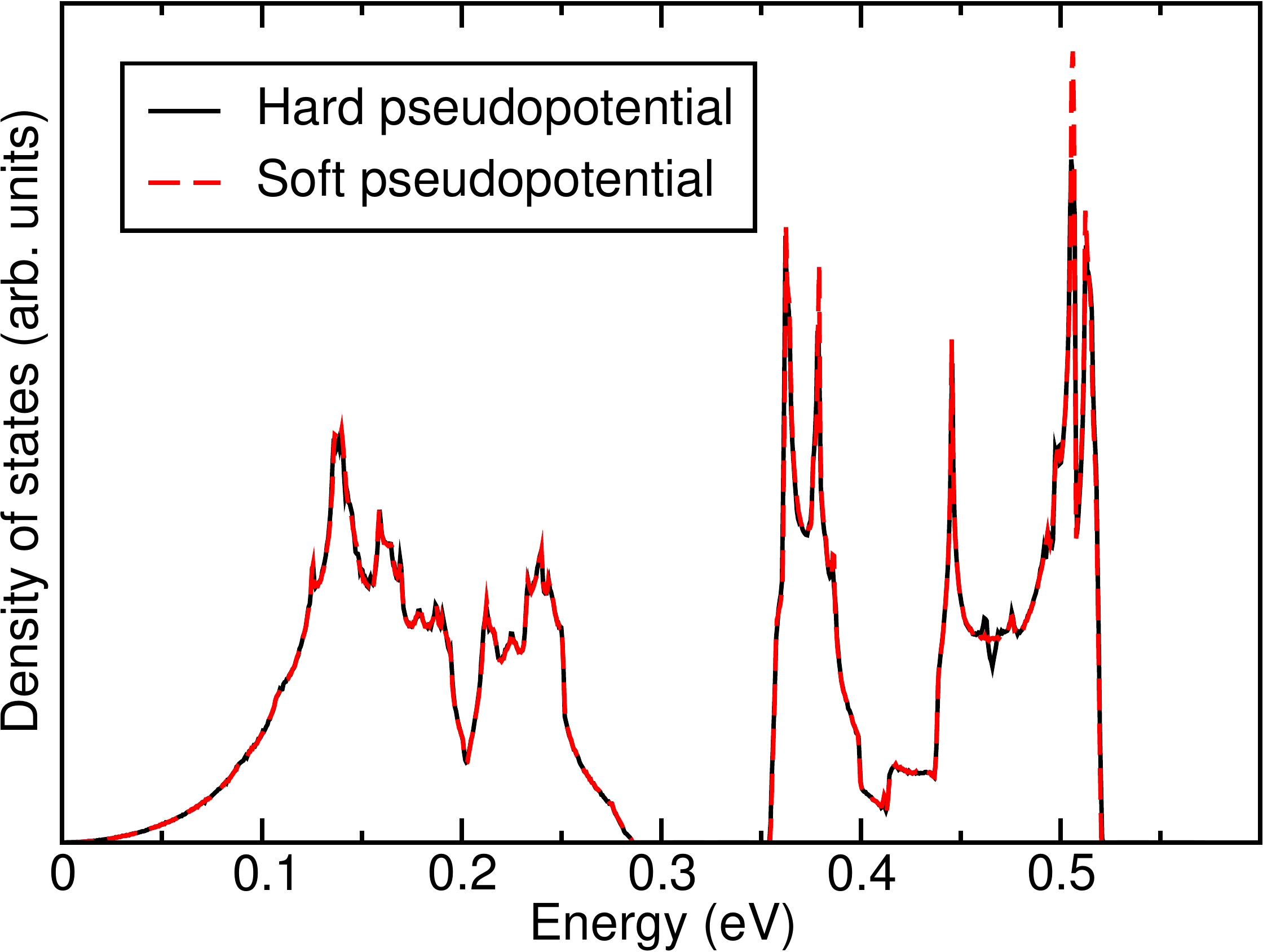}
  \caption{Phonon vibrational density of states for FeHe at $10$\,TPa using a vibrational BZ grid of size $3\times3\times3$ and Fourier interpolation.}
      \label{fig:pseudo}
\end{figure}

We investigate superionicity and melting in the FeHe $Cmcm$ structure using molecular dynamics simulations performed with {\sc quantum espresso}~\cite{quantum_espresso}. We used pseudopotentials converted from {\sc castep} with the following strings:
\begin{verbatim}
He 1|0.7|27|31|36|10(qc=9.2)
Fe 3|1.2|30|35|40|30U:40:31:32(qc=10)
\end{verbatim}
We note that the computational expense of the molecular dynamics calculations forced us to use pseudopotentials with larger core radii than those used for the lattice dynamics calculations. Nonetheless, we have calculated the phonon density of states at $10$~TPa for the FeHe $Cmcm$ structure and confirmed that the hard and soft pseudopotentials lead to consistent results (see Fig.~\ref{fig:pseudo}). We used an energy cutoff of $90$~Ry ($1225$~eV), and a charge density cutoff of $720$~Ry. The electronic Brillouin zone was sampled using a $2\times2\times2$ $\mathbf{k}$-point grid for a $4\times2\times4$ supercell of the conventional cell of the $Cmcm$ structure, containing a total of $256$ atoms.

\begin{figure}[h]
  \includegraphics[scale=0.36]{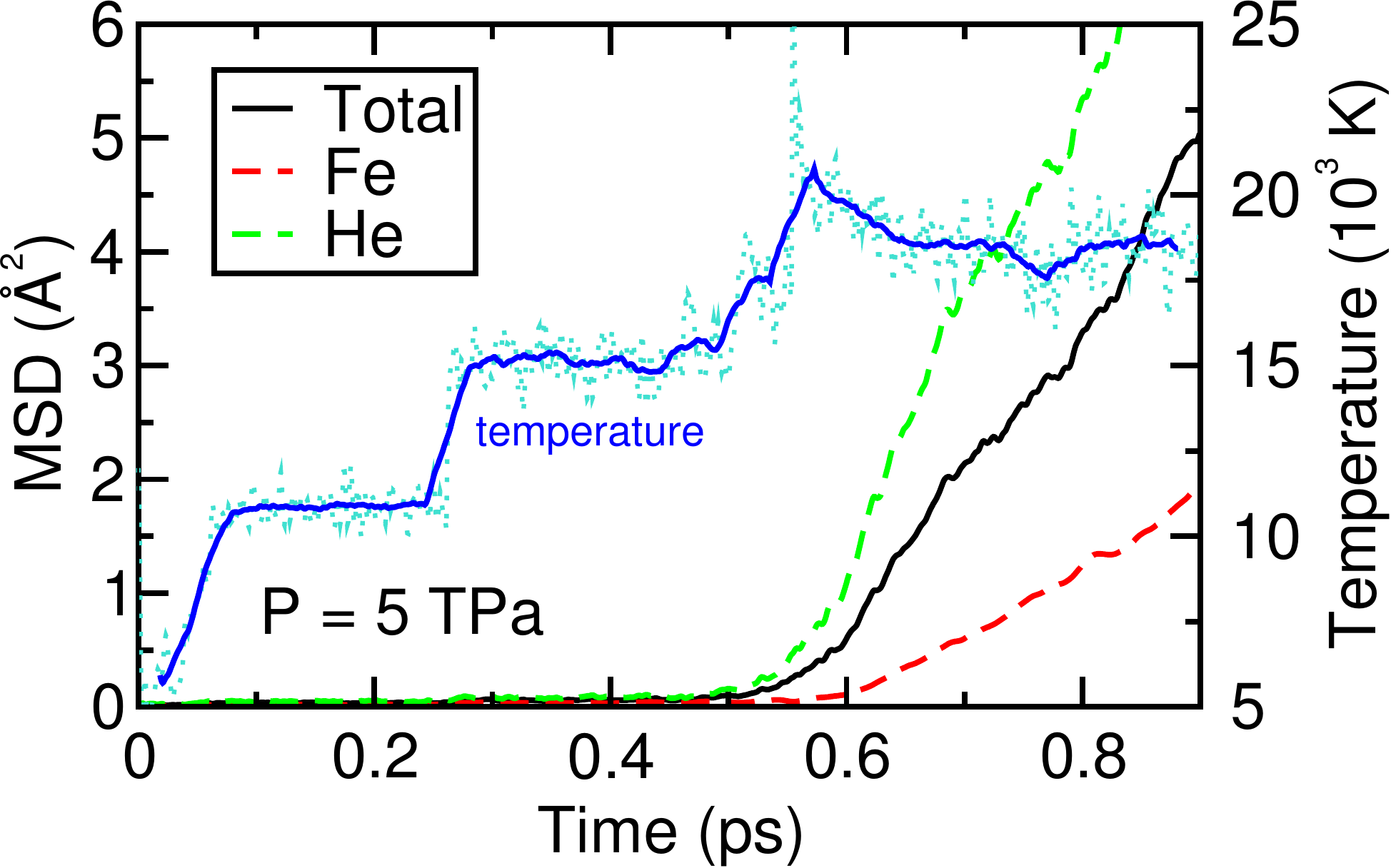}
  \hspace{0.5cm}
  \includegraphics[scale=0.36]{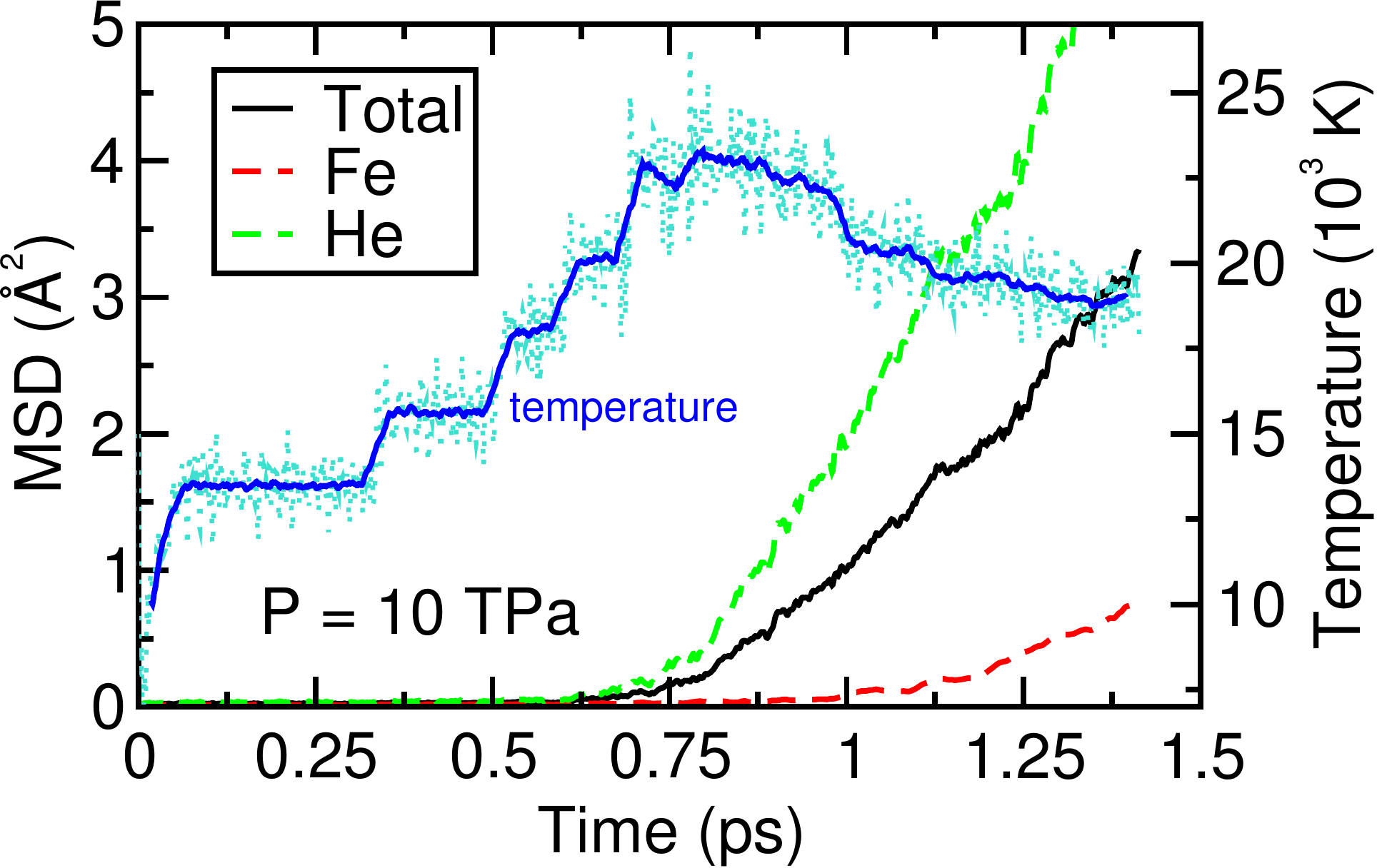}
  \caption{Molecular dynamics runs of $Cmcm$ FeHe. The MSD for the individual Fe and He atoms, as well as the combined MSD are shown on the left vertical axis. The temperature is shown on the right vertical axis, with the blue solid line representing the average over $50$ points, and the light blue dotted line the instantaneous temperature.}
      \label{fig:md}
\end{figure}

The molecular dynamics simulations were used to investigate the melting curve of the FeHe $Cmcm$ compound, for which the Z-method~\cite{melting_z_method} was used. In this method, the temperature is increased to the desired value using an NVT ensemble, and then relaxed using an NVE ensemble. This procedure is iterated while checking whether a component (He or Fe) is diffusing or the whole lattice has melted by monitoring both the temperature and the mean square displacement (MSD) of the relevant species. Representative runs at $5$ and $10$~TPa are shown in Fig.~\ref{fig:md}.

\begin{figure}[h]
  \includegraphics[scale=0.36]{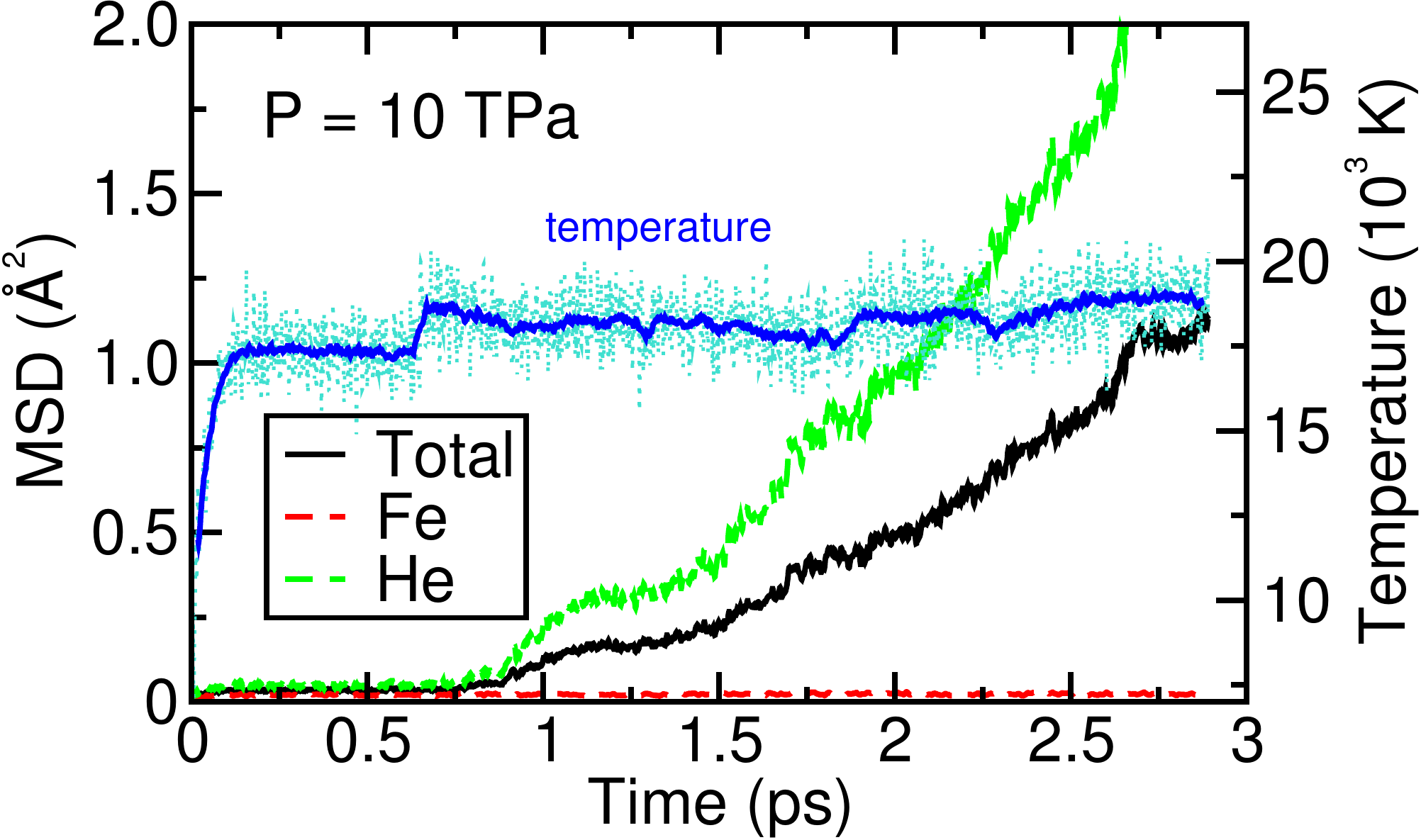}
  \hspace{0.5cm}
  \includegraphics[scale=0.36]{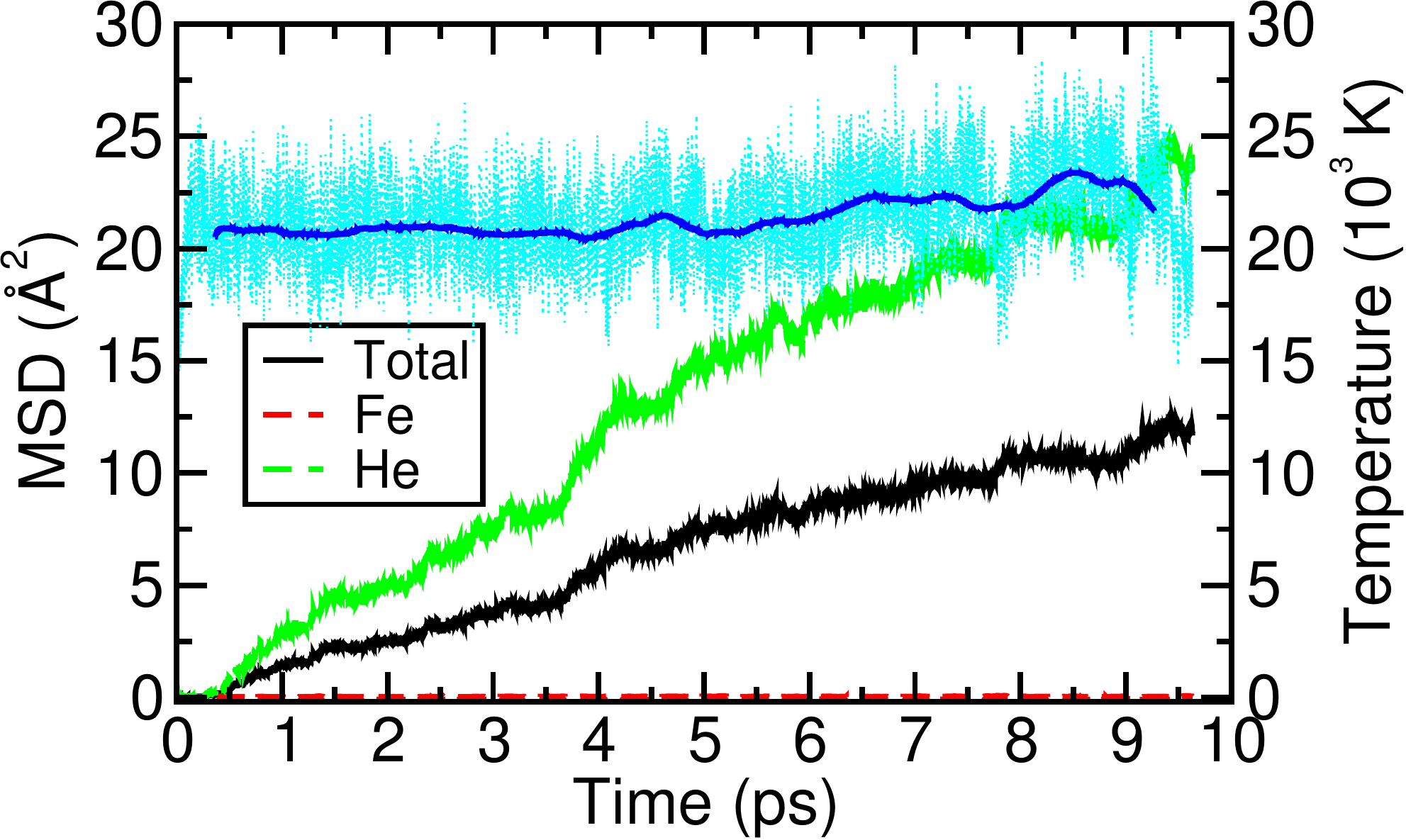}
  \caption{Molecular dynamics runs of $Cmcm$ FeHe in which the helium sublattice is melted but the iron sublattice is not, demonstrating the existence of a superionic phase in FeHe.}
      \label{fig:superionic}
\end{figure}

The results in Fig.~\ref{fig:superionic} are used to prove that the iron sublattice can withstand helium diffusion and therefore sustain the superionic phase. We first equilibrated FeHe at $10$~TPa and $17,500$~K. Next, we equilibrated at $19,000$~K, a temperature at which helium melts and the relaxed temperature then decreases to $18,500$~K. The iron sublattice survives for more than $1.5$~ps, at which point we stop the simulation. We have performed additional tests in which the iron sublattice remains crystalline up to $10$~ps. These molecular dynamics runs provide strong evidence for the existence of the superionic phase in FeHe.

It is interesting to note that the Lindemann criterion correctly predicts the existence of a superionic phase. This criterion associates nuclear vibrations with structural changes and states that a solid melts when the ratio of the root-mean-square nuclear vibrational amplitude and the interatomic separation exceeds some critical value, $\sqrt{\langle u^2\rangle}/a\gtrsim\lambda_{\mathrm{c}}$, which is usually taken to be $\lambda_{\mathrm{c}}=0.1$. The Lindemann criterion using the smallest He-He distance as the interatomic separation $a$ predicts that the helium sublattice melts at temperatures of a few thousand Kelvin, but at those temperatures the Fe-Fe Lindemann ratio is significantly smaller than $\lambda_{\mathrm{c}}$. These results suggest that, upon increasing temperature, the helium chains melt within the iron channels in FeHe before the iron channels themselves melt. We note that quantitatively, the Lindemann criterion results significantly deviate from the molecular dynamics results.

\section{Helium and iron melting}

We have calculated the helium melting curve using MD in a similar manner to that described above for FeHe $Cmcm$. We will report full details of this calculation in a separate publication.

The iron melting curve is taken from Ref.~\cite{iron_melting_dft_tpa}, in which density functional theory in conjunction with a two-phase coexistence approach was used.

\end{document}